\documentclass[10pt,a4paper]{article} 
\usepackage{multicol}  
\usepackage{float} 
\usepackage{booktabs} 
\usepackage{article_styles}
\usepackage{pdflscape}
\usepackage[english]{babel}  
\usepackage[hyphens]{url}
\usepackage[hidelinks]{hyperref}
\hypersetup{breaklinks=true}
\usepackage{changes}
\usepackage{todonotes}
\usepackage{comment}

\usepackage[numbers]{natbib}

\begin{document}

\title{Prediction of healthcare costs on consumer direct health plan in the Brazilian context}

\author[1,a]{Peixoto, Claudia M.}
\author[1,b]{Marcondes, Diego}
\author[1,c]{Melo, Mariana P.}
\author[2,d]{Maia, Ana C.}
\author[1,e]{Correia, Luis A.}
\affil[1]{Instituto de Matemática e Estatística, Universidade de São Paulo, Brazil}
\affil[2]{Departamento de Contabilidade e Atuária, Universidade de São Paulo, Brazil}
\affil[a]{claudiap@ime.usp.br}
\affil[b]{dmarcondes@ime.usp.br}
\affil[c]{marianapm@usp.br}
\affil[d]{anacmaia@usp.br}
\affil[e]{luis.correia@ime.usp.br}

\maketitle

\thispagestyle{fancy}

\begin{minipage}{\linewidth}
\begin{abstract}
The rise in healthcare costs has led to the adoption of cost-sharing devices in health plans. This article explores this discussion by simulating Health Savings Accounts (HSAs) to cover medical and hospital expenses, supported by catastrophic insurance. Simulating 10 million lives, we evaluate the utilization of catastrophic insurance and the balances of HSAs at the end of working life. To estimate annual expenditures, a Markov Chains approach - distinct from the usual ones - was used based on recent past expenditures, age range, and gender. The results suggest that HSAs do not create inequalities, offering a viable method to sustain private healthcare financing for the elderly.
\end{abstract}

\end{minipage}

\section*{Keywords}

health savings accounts, consumer directed health plans, prediction of healthcare expenses, Markov Chain

\clearpage

\begin{multicols}{2}  

\flushbottom

\selectlanguage{english}  
\thispagestyle{empty}

\section{Introduction}

Consumer directed health plans (CDHP) are products intended to fund healthcare expenses, which have been consolidated in the United States since the end of the nineties as a possible approach to contain the rise in healthcare costs. In CDHP, the insured has a personal account to pay for his or her medical procedures, what leads to the increase in awareness about healthcare costs (\cite{gabel_lasso2002} and \cite{Bundorf2016}).

According to \cite{Bundorf2016}, there are three features often associated to CDHP: relatively high deductibles, a personal account to accumulate resources, and availability of information about healthcare costs. On the one hand, these features diverge from models based purely on mutualism, as they seek to accomplish a higher perception of well-being and economy efficiency, while mitigating the disruptive impacts due to moral hazard. On the other hand, the attachment of individual expenses to a personal account raises questions about justice, as individuals could persistently experience health shocks during their work life, making it unlikely for them to save enough to cover their expenses after retirement. This would cause, from a population point of view, a great variation on the balance of these accounts, causing a part of the population to have a great amount of resources at retirement (savings), while other part would not have accumulated anything (the product would have features of a self-insurance)\cite{Eichner1998}.

In recent years, a great rise in healthcare costs has been observed around the world. In the Brazilian supplementary health system, for example, the mean annual healthcare expense, including dental benefits, increased 73\% between 2013 and 2018 according to the National Supplementary Health Agency (ANS)\cite{ANS20192021:dados}, what has led to a public debate about the sustainability of the sector\cite{ANS20192021:mapa}. Following the example of other countries, it has been proposed an alternative to the system in an attempt to establish norms and legal certainty for products with co-payment and deductible devices, as it is a reasonable assumption that Brazil faces similar challenges as other countries in what concerns the essential cause of healthcare costs increase, that is, spending more than needed to treat a specific health problem. This over-consumption and oversupply pattern is understood as the result of two factors. On the one hand, consumers may partake in more risky activities when protected by a health plan, increasing the probability of needing more care (ex-ante moral hazard) or simply consuming more care than necessary (ex-post moral hazard). On the other hand, providers may supply more care than needed, inducing the demand (\cite{Ehrlich_Becker1972} and \cite{Zweifel_Manning2000}).

In order to mitigate the impact of these factors, in 2018 the Brazilian regulatory agency (ANS) proposed a resolution to regulate co-payment and deductibles in health plans (RN nº 433, June 27 2018), which had a negative public repercussion, that led ANS to suspend the norms, leaving the market uncovered in what concerns the understanding of these devices. The negative repercussion was caused by a misunderstanding about the real use of these devices in health plans, what evidences the need to further the discussion about them, including the possibility of considering them as part of a new market segment.

This paper aims to subsidize the discussion with the analysis of empirical data by simulating health savings accounts (HSA). A product derived from CDHP, the HSA are individual accounts set to exclusively cover expenses with healthcare goods and services, which is composed by a resources accumulation phase, typically subsidized by the employer with participation of the employee, followed by a decumulation phase, generally after retirement. The objective of this form of funding is the implementation of a risk pooling over the individual cycle of life, mobilizing resources from various sources to fund the decumulation phase.

In the literature, several studies (\cite{Arrow1963}, \cite{Pauly1968} and \cite{Zeckhauser1970}) emphasize that the theoretical perspective of HSA addresses incentives leading to moral hazard, a significant inefficiency in the healthcare goods and services market. This moral hazard has been identified as one of the primary causes of cost inefficiencies in the sector \cite{Goodman_Musgrave1992}. Furthermore, in mobilizing resources to a savings account for the stages of the cycle of life when the increase in healthcare expenses extrapolate that in income, the HSA are an answer to policies which aim to maintain private funding of healthcare for the elderly population.

Since the end of the nineties, countries as the United States, China, Singapore and South Africa have taken steps to incorporate these alternatives in their health systems. In the US experience, the HSA products combine a high deductible health plan with tax breaks to form a savings account. These products have a mixed framework, which seeks to reduce the individual exposition to the risk of extreme events, by combining two devices: a catastrophic health insurance associated with a personal savings account.

In this paper, we simulate a HSA product combined with a catastrophic health insurance during a labor period of forty years (from 25 to 65 years-old) in the Brazilian context. The main objective of the simulation is to study, from a population point of view, the variation of HSA balances at the end of the accumulation phase. A great discrepancy in these balances may be interpreted as the result of persistent health shocks, understood here as the cause of the financial need for healthcare goods and services, and measured through the costs covered by a health plan under the individual perspective. The persistence of shocks is a fundamental aspect in the debate about CDHP since one could argue that individuals with poor health would not be able to accumulate savings over time, so these accounts would reflect inequalities between individuals with distinct overall health conditions.

In order to simulate the annual healthcare expenses of an individual, we propose an approach based on Markov Chains \cite{Taylor_Karlin1998}, which differs from the usual methods of predicting healthcare expenses based on regression models \cite{Jones_2000}. In this approach, we define levels of annual expense, which are ranges of expense, and apply a prediction technique which estimates the probability of an individual to have annual expenses in each level, based on his or her levels in the previous two years, age range and sex. This model does not try to predict the exact value of the annual expenses by assuming some kind of functional relation between the expenses and the independent variables, but rather predicts the probability of each level of expense for each combination of categories of the independent variables sex, age range and previous expense levels.

Freed from assumptions about regression error distributions and the restricted form of the functional relation between dependent and independent variables, which are often not satisfied in real datasets, a Markov Chain approach makes a mild assumption about the dependence between the expenses in a year with that in previous years, and tries to achieve a more realistic goal, which is to predict an expense level rather than the exact value of the expense. This more realistic approach may lead to improved prediction models, as it can be applied in the context of any country and naturally considers their specific behaviors.

In Section 2 we present the dataset used in the simulation, and discuss the considered CDHP product, the modelling of healthcare expenses and the proposed Markov Chain approach. In Section 3 we present the results of the simulation studies, and in Section 4 we comment on how they may be interpreted to subsidize the debate about CDHP products in Brazil.

\section{Materials and Methods}

\subsection{Dataset}

The dataset contains claims of a Brazilian self-management health plan. There are six types of health plans operators in Brazil, which are characterized by the juridical nature of their operations. The self-management health plans are similar to the self-insured group health plans in the US, which offer plans for employees and dependents, assuming the financial risk, which can be partially funded by the employees. This portfolio was followed longitudinally for five years (2005 to 2009) and all claims (expenses) are corrected by the inflation index IPCA (Extended National Consumer Price Index) to the December 2009 value.

As the objective of this paper is to study the persistence of health shocks during the work life, we will consider only working age individuals, that is, those with 25 years completed at January 2005 and with at most 65 years completed at December 2009, which amounts to approximately 39,000 lives. During the five years of study, the portfolio size changed because of the entrance of new individuals, death and other motives, so from all these lives, around 11,000 were not followed during all the period, so there will be considered in the analysis only the 27,780 individuals between 25 and 65 years-old which were in the portfolio for the whole period.

Table 1 presents some descriptive statistics of the annual expenses of the individuals within the considered age range which stayed in the portfolio during the five-year period. The percentage of individuals with zero annual expenses is between 5 and 6\% in all years. As Table 1 portrays the behavior over time of a closed cohort, it is expected that a shift in the mean expense occurs due to the aging of the group. However, even though the variation in the mean expense from 2005 to 2009 was around 36\% and the expenses are corrected by a general inflation index, there is no guarantee that the real variation due to aging can still be observed in the costs of the healthcare goods and services of this portfolio. Therefore, in reality, both the aging effect and the real variation in the costs are reflected in these values.

As expected (\cite{Seshamani_Gray2004}, \cite{Zweifel_etal2004} and \cite{Werblow_etal2007}), large expenses are concentrated in a small parcel of individuals: in 2008, lesser than 1\% of the individuals were responsible for expenses ranging from R\$ 31,158 to R\$ 1,044,525 where we observe that the maximum expense is 385 times the mean one. Also, in 2005, around 5\% of the individuals were responsible for annual expenses ranging from R\$ 7,083 to R\$ 426,772, and the maximum expense was 206 times the mean one.

In order to compare men and women according to their annual expenses, we present in Tables 2.1 and 2.2 in Appendix A some descriptive statistics for the annual expenses for each sex. From the total of individuals, there are 13,539 (49\%) women and 14,241 (51\%) men. Over time, the percentage of women without expenses ranged between 5.1 and 5.5\%, while the percentage of men ranged from 5.8 to 6.8\%. When considering only the individuals with positive expense, we see that the mean annual expenses of women is greater than that of men, as it varied between R\$ 2,341 and R\$ 3,062 for the female sex, and between R\$ 1,795 and R\$ 2,576 for the male sex. In the same manner, the median annual expenses of women ranged from R\$ 998 to R\$ 1,104 and that of men from R\$ 580 to R\$ 721.

\end{multicols} 

% latex table generated in R 4 2.3 by append_kable_to_file function
% Current Date/Time: 2024-02-18 00:57:55
\begingroup
\renewcommand{\thetable}{1}
\begin{table}[H]
\centering
\begin{tabular}[t]{>{}l>{\raggedleft\arraybackslash}p{2.5cm}>{\raggedleft\arraybackslash}p{2.5cm}>{\raggedleft\arraybackslash}p{2.5cm}>{\raggedleft\arraybackslash}p{2.5cm}>{\raggedleft\arraybackslash}p{2.5cm}}
\toprule
Description & 2005 & 2006 & 2007 & 2008 & 2009\\
\midrule
\textbf{\cellcolor{gray!10}{n}} & \cellcolor{gray!10}{27,780} & \cellcolor{gray!10}{27,780} & \cellcolor{gray!10}{27,780} & \cellcolor{gray!10}{27,780} & \cellcolor{gray!10}{27,780}\\
\textbf{PctNoExpense} & 5.97 & 5.52 & 5.77 & 5.55 & 5.76\\
\textbf{\cellcolor{gray!10}{p25}} & \cellcolor{gray!10}{319} & \cellcolor{gray!10}{303} & \cellcolor{gray!10}{332} & \cellcolor{gray!10}{368} & \cellcolor{gray!10}{359}\\
\textbf{p50} & 764 & 736 & 791 & 904 & 872\\
\textbf{\cellcolor{gray!10}{p75}} & \cellcolor{gray!10}{1,746} & \cellcolor{gray!10}{1,666} & \cellcolor{gray!10}{1,789} & \cellcolor{gray!10}{2,090} & \cellcolor{gray!10}{2,008}\\
\addlinespace
\textbf{p90} & 3,993 & 3,735 & 4,130 & 4,861 & 4,761\\
\textbf{\cellcolor{gray!10}{p95}} & \cellcolor{gray!10}{7,083} & \cellcolor{gray!10}{6,531} & \cellcolor{gray!10}{7,477} & \cellcolor{gray!10}{8,726} & \cellcolor{gray!10}{9,114}\\
\textbf{p96} & 8,497 & 7,960 & 8,864 & 10,247 & 10,780\\
\textbf{\cellcolor{gray!10}{p97}} & \cellcolor{gray!10}{10,450} & \cellcolor{gray!10}{9,853} & \cellcolor{gray!10}{10,935} & \cellcolor{gray!10}{12,910} & \cellcolor{gray!10}{14,190}\\
\textbf{p98} & 13,538 & 13,134 & 15,380 & 18,237 & 20,160\\
\addlinespace
\textbf{\cellcolor{gray!10}{p99}} & \cellcolor{gray!10}{22,910} & \cellcolor{gray!10}{21,683} & \cellcolor{gray!10}{25,851} & \cellcolor{gray!10}{31,158} & \cellcolor{gray!10}{33,730}\\
\textbf{p995} & 33,905 & 35,695 & 43,747 & 54,407 & 55,601\\
\textbf{\cellcolor{gray!10}{p999}} & \cellcolor{gray!10}{92,962} & \cellcolor{gray!10}{105,585} & \cellcolor{gray!10}{92,549} & \cellcolor{gray!10}{139,811} & \cellcolor{gray!10}{181,947}\\
\textbf{max} & 426,772 & 528,293 & 355,220 & 1,044,525 & 965,287\\
\textbf{\cellcolor{gray!10}{mean}} & \cellcolor{gray!10}{2,064} & \cellcolor{gray!10}{2,041} & \cellcolor{gray!10}{2,220} & \cellcolor{gray!10}{2,710} & \cellcolor{gray!10}{2,813}\\
\addlinespace
\textbf{sd} & 6,833 & 8,127 & 7,536 & 11,612 & 12,673\\
\bottomrule
\end{tabular}
\label{tab:table1}
\end{table}
\endgroup
\textbf{n}: total of individuals in portfolio;\textbf{p(k)}: k-th sample percentile;\textbf{SD}: standard deviation

\medskip

\textbf{Table 1}. Descriptive statistics of the annual expenses of individuals between 25 and 65 years-old which stayed on the portfolio during the whole five-year period. The percentiles, mean and standard deviation are calculated considering only the individuals with positive expenses.

\begin{multicols}{2}  

In Tables \ref{tab:table3.1} to \ref{tab:table3.18} in Appendix A we present the descriptive statistics of the annual expenses by sex and age range, where it can be seen in all years that the mean annual expenses increase with age, highlighting the effect of age range on healthcare expenses. When we assess the factors which influence individual healthcare expenses, age is always presumed to have a positive effect since, as age increases, so does the probability of occurrence of chronicle diseases and loss of functional capabilities. Therefore, it is expected that high expenses are related to advanced age (\cite{Duncan_etal2016} and \cite{Frees_etal2014}).

In Figure \ref{fig:ci2005a2009_facet} in Appendix A we present the mean annual expenses for each combination of sex and age range, with an error bar representing one standard error. On top of each error bar, we present the size of each group. When we compare men and women, we see a distinct pattern, as women tend to spend more in the early age ranges, a tendency which inverts itself in later ranges. This feature is found in other references in the literature \cite{Yamamoto_2013}. Over time, we observe a greater increment in the expenses of the later age ranges.

While the mean annual expenses of men increase constantly with age, the mean annual expenses of women is stable until the age of 50, when we see an increase in the mean of the age ranges 51-55 and 56-60, followed by a small reduction in age range 61-65. Thus, we see that women spend more than men in practically all age ranges, although this difference decreases with age. In 2006 and 2008, the mean annual expenses of men in the age range 61-65 surpassed that of women.

\subsection{Persistence of costs}

Persistence of costs is defined as the continuity (permanence) of elevated health costs of an individual over time. In order to analyze the persistence of costs, we divide the individual annual expenses of each year $i \in \{2005, 2006, 2007, 2008, 2009\}$ into four levels, namely:

\begin{itemize}
    \item $F_{1,i}$: annual expenses lesser or equal to R\$ 300 at year \textit{i};
    \item $F_{2,i}$: annual expenses greater than R\$ 300 and lesser or equal to R\$ 1,000 at year \textit{i};
    \item $F_{3,i}$: annual expenses greater than R\$ 1,000 and lesser or equal to R\$ 5,000 at year \textit{i}; and
    \item $F_{4,i}$: annual expenses greater than R\$ 5,000 at year \textit{i}.    
\end{itemize}

Dividing the expenses this way we have, for each individual, a sequence of five levels describing his or her expenses throughout the years. For example, an individual with costs $\{0; 200; 500; 1,500; 200\}$ has sequence $F_{1,2005},F_{1,2006},F_{2,2007},F_{3,2008},F_{1,2009}$ as expense levels.

\end{multicols}

\begin{figure}[H]
    \centering
    \includegraphics[scale=1.0]{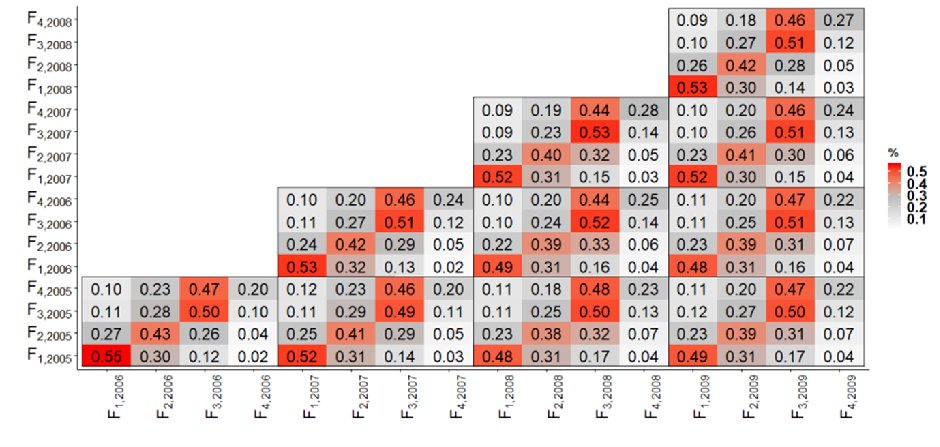}
    \caption{Transition matrix of expenses for each two years combination.}
    \label{fig:trmat_fig2}  % 
\end{figure}

\begin{multicols}{2}  

In Figure \ref{fig:trmat_fig2} we present the sample proportion of individuals which transitioned between each pair $(F_{k,i}, F_{l,j}), j > 1$ of states, in which rows represent the origin \textit{state} $F_{k,i}$ and columns the destination \textit{state} $F_{l,j}$. The color of the matrices entries is related to the estimated transition probability: red means great probability (around 0.5), gray means medium probability (around 0.2) and white means low probability (around zero). In this figure, there are ten 4 by 4 matrices, one for each pair of distinct years, which present the transition probabilities from the ranges of the early year to that of the later. For example, the matrix in the lower-left corner presents the transition probabilities between the expense’s levels of 2005 and 2006, and the matrix in the upper between the expense’s levels of 2008 and 2009. The leading diagonal of each matrix is the proportion of individuals which stayed at the same expense level at both years.

In Figure \ref{fig:trmat_fig2} we see, for example, that within individuals with expenses greater than R\$ 5,000 in 2008, 46\% had expenses between R\$ 1,000 and R\$ 5,000, and 27\% maintained the expenses at level greater than R\$ 5,000, in 2009; within individuals with annual expenses lesser or equal to R\$ 300 in 2007, 52\% had expenses lesser than R\$ 300, and only 3\% had expenses greater than R\$ 5,000, in 2008. We also note that high values of probabilities are concentrated around the leading diagonals, except at the greatest expense level. However, these high values decrease with the distance between the years, highlighting the fact, also observed in Eichner, Mcclellan \& Wise \cite{Eichner_etal1996}, that persistence of costs decays over time. It is important to note that the matrices were constructed considering only the individuals between 25 and 65 years-old that were alive in 2009. If the individuals which died in the considered period were also considered, then there would be greater probabilities in the diagonal point of the greatest expense level, as it is known that individuals tend to have consistently high expenses in the period leading up to their death.

In order to evaluate the effect of age on the persistence of costs, we present in Fig. \ref{fig:trmat_fig3_4} in Appendix A the transition matrices calculated considering only the younger (21 to 40 years-old) and older (41 to 65 years-old) individuals, respectively. When comparing the diagonal of the matrices of both figures, we see that, for younger individuals, the probability of those in expense level 3 remaining in the same level or moving up to expense level 4 is approximately 56\%. However, for the population aged 40 and above, this proportion is around 66\%, i.e., 10 percentage points higher, in average.

\subsection{Health Savings Accounts (HSA)}

The health savings accounts (HSA) are devices in which a savings account attached to each individual receives annual contributions with the exclusive purpose of covering healthcare expenses. The annual expenses covered by HSA are limited by a predefined value while its balance is positive. In case expenses exceed the limit or the account balance, then a catastrophic insurance is activated to cover them. Although there is a limitation on the expenses covered by funds from the savings account, in this device there is no limitation on individual annual healthcare expenses, as the insurance covers any expenses beyond the threshold.

The CDHP product considered in this paper is a HSA in which each individual has an account, started at 25 years-old, from which are deducted his or her healthcare expenses. The account dynamic is as follows:

\begin{itemize}
    \item At the beginning of each year, the employer deposits R\$ 2,500\footnote{[Include the rationale of choosing this amount as reference for HSA deposits – w/ Ana]} on the individual account.
    \item In case the annual healthcare expenses of an individual do not exceed the account balance or R\$ 5,000, they are fully paid by funds from the account. If the annual expenses surpass R\$ 5,000 or the balance of the account, then R\$ 5,000, or the balance, is withdrawn to partially cover them, and the remaining value is covered by a catastrophic insurance. Therefore, all healthcare expenses up to R\$ 5,000 or the account balance, if lesser, are covered by the individual, and the remaining value is covered by the insurance.
\end{itemize}

In what follows, we present an approach based on Markov Chain to simulate the product described above, which seeks to predict the expense level of each individual at each year of his or her work life in order to assess the balance of the accounts at age 65, and the frequency and severity of catastrophic insurance use, from a population point of view. In the next section, we discuss the approaches to healthcare expenses prediction at the individual level used in the literature, presenting their qualities and shortcomings.

\subsection{Prediction of future healthcare expenses at individual level}

Modelling techniques to assess the risk associated with events which incur in healthcare expenses are relatively recent \cite{Jones_2000}, for their development is dependent on the volume and quality of available information, which have only improved in the last couple of decades. Also, the sector has been demanding new frameworks for covering and management of health risks, in contrast to its historical foundation, based purely on refunding, increasing the need for improved quantitative models.

In order to model healthcare expenses patterns in the US, Frees, Gao \& Rosenberg \cite{Frees_etal2011} considered more than thirty factors related to an individual's demography, socioeconomic level, health condition, employment and availability of health insurance. There were considered two dependent variables, namely, frequency and severity of expenses, and the significant factors found for each of them differed. Also, Duncan, Loginov \& Ludkovski \cite{Duncan_etal2016} explored the effects on healthcare expenses of approximately one hundred independent variables related to demography, age and comorbidities history, among others.

Pope et al  \cite{Pope_etal2004} and Duncan, Loginov \& Ludkovski \cite{Duncan_etal2016} pointed out the limitations of models which consider only demographic factors and health condition history, affirming the importance of taking into account past expenses while developing prediction models. Indeed, according to Duncan, Loginov \& Ludkovski \cite{Duncan_etal2016}, one of the best sources of information for predicting healthcare expenses are past expenses.

About issues encountered when modelling individual healthcare expenses, Duncan, Loginov \& Ludkovski \cite{Duncan_etal2016} evidenced that: (a) annual expenses follow approximately a log-normal distribution and are characterized by high variance; (b) the error distribution is clearly heteroscedastic: the conditional variance for individuals with high expenses is greater than that of individuals with low expenses; (c) there exists interaction between the independent variables, and the relationship between expenses and health condition is expected to be non-linear, what implies non-linear relationships between dependent and independent variables; (d) many of the independent variables are highly correlated and some of them have very rare categories. Due to the great variability of annual expenses in a population and the great number of independent variables, the quality of fitted models is quite low.

Other problems encountered when modelling annual healthcare expenses are the excess of zeros and the presence of extreme values. The first problem is due to individuals which do not use their health insurance in the period of a year. The second is caused by the fact that a very small portion of a population has annual expenses hundreds of times the mean one. In order to attack these problems, Marcondes, Peixoto \& Maia \cite{Marcondes_etal2018} proposed a hurdle model for heavy-tailed data, which is specifically useful for annual healthcare expenses data. A hurdle model takes special care of the zero values, while a heavy-tail model takes into account values far from the mean. The mixture of models treating these two features is a better fit for annual healthcare expenses than the usual regression models, as evidenced by the application in Marcondes, Peixoto \& Maia \cite{Marcondes_etal2018}, which considered a dataset of annual healthcare expenses.

Based on the literature, we see that there is a need to develop better techniques to model healthcare expenses in a dynamic scenario characterized by the rapid increase in the amount of available information and data about individuals. With this purpose, we consider a Markov Chain approach to model healthcare expenses in order to simulate an HSA product.

\subsection{Methodology for prediction of future healthcare expenses at individual level}

In this section, we present the methodology proposed in this paper to predict future healthcare expenses at the individual level, which is based on Markov Chains of order 2 that are defined in Appendix B. The prediction technique will be employed to estimate annual individual healthcare expenses in order to simulate the expenses which will be covered by a HSA during the labor life of simulated individuals.

\subsubsection{Simulation of individual annual healthcare expenses over time}\label{sec:markov}

Our simulation study is based on the assumption that the pattern of annual healthcare expenses of an individual is similar to that of those of the same sex, age range and expenses history. The study consists in the simulation of 10,000 lives, starting at 25 years-old, which are followed annually until 65 years-old, what corresponds to the forty-one years of a work life. The annual expenses of these individuals in their first year (25 years-old) are sampled from the 2009 annual expenses empirical distribution of age range 25-30. There are 1,683 individuals in the age range 25-30 in 2009, so in order to obtain 10,000 lives from them, we will sample 10,000 lives with repetition from the 1,683 individuals, setting the age of all new sampled individuals to exactly 25 years-old.

From the sex, age range, expenses history and expenses in the first year of the 10,000 lives, we simulate their personal HSA over time, a process fulfilled by iterating two consecutive procedures. In the first step, we predict to which expense level each individual will go, based on probabilities depending on his or her sex, age range at the present year and expense level on the current and previous year. On the second step, we sample the exact value to be deducted from each individual account to cover the expenses from the empirical distribution of the annual expenses of the individuals of the same sex, age range and predicted expense level. This empirical distribution is the one in Fig. \ref{fig:empdistlbl} which is related to the individual's age range and sex. The distribution is built by combining individual's medical expenses from years $2005$ to $2009$, stratified by age and sex, obtaining the possible points that can be sampled within the expense level sampled in the first step, i.e., between the respective dotted lines in Fig. \ref{fig:empdistlbl}. The simulation was replicated $1,000$ times. In Appendix B, we further explain the simulation dynamics.

\section{Results}

In order to assess the effectiveness of the HSA, we present in this section the results of the simulation study regarding the balance of the individual accounts over time, and the frequency and severity of the catastrophic insurance use.

\subsection{Balance of the individual health savings accounts}

In Table 5 we present descriptive statistics of the individual HSA of the simulated 10,000 lives at every five years between 25 and 65 years-old.

\medskip

\end{multicols}

% latex table generated in R 4 2.3 by append_kable_to_file function
% Current Date/Time: 2024-02-24 20:06:10
\begingroup
\begin{table}[H]
\centering
\fontsize{7}{9}\selectfont
\begin{tabular}[t]{>{\raggedright\arraybackslash}p{0.65cm}>{\raggedleft\arraybackslash}p{0.55cm}>{\centering\arraybackslash}p{0.55cm}>{\raggedleft\arraybackslash}p{0.55cm}>{\centering\arraybackslash}p{0.55cm}>{\raggedleft\arraybackslash}p{0.55cm}>{\centering\arraybackslash}p{0.55cm}>{\raggedleft\arraybackslash}p{0.55cm}>{\centering\arraybackslash}p{0.55cm}>{\raggedleft\arraybackslash}p{0.55cm}>{\centering\arraybackslash}p{0.55cm}>{\raggedleft\arraybackslash}p{0.55cm}>{\centering\arraybackslash}p{0.55cm}>{\raggedleft\arraybackslash}p{0.55cm}>{\centering\arraybackslash}p{0.55cm}>{\raggedleft\arraybackslash}p{0.55cm}c}
\toprule
\multicolumn{1}{c}{ } & \multicolumn{2}{c}{25-30} & \multicolumn{2}{c}{31-35} & \multicolumn{2}{c}{36-40} & \multicolumn{2}{c}{41-45} & \multicolumn{2}{c}{46-50} & \multicolumn{2}{c}{51-55} & \multicolumn{2}{c}{56-60} & \multicolumn{2}{c}{61-65} \\
\cmidrule(l{3pt}r{3pt}){2-3} \cmidrule(l{3pt}r{3pt}){4-5} \cmidrule(l{3pt}r{3pt}){6-7} \cmidrule(l{3pt}r{3pt}){8-9} \cmidrule(l{3pt}r{3pt}){10-11} \cmidrule(l{3pt}r{3pt}){12-13} \cmidrule(l{3pt}r{3pt}){14-15} \cmidrule(l{3pt}r{3pt}){16-17}
  & $\mu$ & $\sigma$ & $\mu$ & $\sigma$ & $\mu$ & $\sigma$ & $\mu$ & $\sigma$ & $\mu$ & $\sigma$ & $\mu$ & $\sigma$ & $\mu$ & $\sigma$ & $\mu$ & $\sigma$\\
\midrule
n0 & 2,693 & 46.38 & 496 & 21.93 & 227 & 14.92 & 144 & 11.69 & 121 & 10.89 & 123 & 10.53 & 126 & 10.61 & 133 & 11.61\\
p5 & 1,058 & 14.09 & 2,907 & 71.85 & 5,138 & 125.04 & 7,434 & 174.39 & 9,238 & 214.90 & 10,232 & 248.86 & 10,734 & 278.85 & 10,706 & 294.34\\
p10 & 1,593 & 16.48 & 4,711 & 72.43 & 7,961 & 122.21 & 11,123 & 166.51 & 13,734 & 202.02 & 15,413 & 237.51 & 16,479 & 263.42 & 16,807 & 280.47\\
p15 & 1,960 & 12.16 & 6,165 & 77.60 & 10,056 & 118.58 & 13,810 & 161.58 & 16,966 & 190.51 & 19,139 & 221.65 & 20,622 & 251.52 & 21,249 & 271.23\\
p25 & 2,385 & 6.84 & 8,467 & 75.27 & 13,298 & 111.78 & 17,930 & 141.20 & 21,876 & 173.91 & 24,800 & 204.01 & 26,927 & 233.74 & 28,062 & 254.14\\
\addlinespace
p40 & 3,993 & 22.48 & 11,214 & 70.06 & 17,117 & 101.10 & 22,734 & 127.61 & 27,598 & 157.55 & 31,423 & 188.78 & 34,341 & 217.03 & 36,126 & 238.85\\
p50 & 4,723 & 14.96 & 12,846 & 65.88 & 19,359 & 95.41 & 25,535 & 123.00 & 30,934 & 150.65 & 35,309 & 179.50 & 38,721 & 206.08 & 40,932 & 231.85\\
p75 & 7,843 & 39.48 & 16,608 & 53.84 & 24,777 & 82.13 & 32,278 & 109.82 & 39,016 & 135.52 & 44,831 & 167.13 & 49,603 & 193.85 & 53,051 & 227.11\\
p85 & 9,404 & 23.11 & 18,587 & 50.79 & 27,201 & 73.21 & 35,361 & 102.83 & 42,753 & 133.39 & 49,286 & 164.03 & 54,800 & 189.26 & 58,949 & 230.20\\
p95 & 11,792 & 22.48 & 21,663 & 56.93 & 30,869 & 76.88 & 39,728 & 101.04 & 48,068 & 132.23 & 55,699 & 168.59 & 62,406 & 209.51 & 67,718 & 253.78\\
\addlinespace
p98 & 13,193 & 29.56 & 23,301 & 46.22 & 32,987 & 77.81 & 42,280 & 103.93 & 51,064 & 140.66 & 59,313 & 181.59 & 66,747 & 238.78 & 72,770 & 291.57\\
max & 14,621 & 50.65 & 26,407 & 114.55 & 37,976 & 194.88 & 49,347 & 318.00 & 60,427 & 526.57 & 71,231 & 762.78 & 81,510 & 1,064.25 & 89,269 & 1,342.32\\
mean & 5,446 & 20.08 & 12,571 & 50.98 & 18,858 & 74.16 & 24,831 & 95.64 & 30,115 & 115.29 & 34,478 & 134.69 & 37,971 & 151.71 & 40,345 & 168.52\\
sd & 3,410 & 8.89 & 5,613 & 24.45 & 7,780 & 39.52 & 9,799 & 53.11 & 11,769 & 65.71 & 13,756 & 79.44 & 15,599 & 91.95 & 17,186 & 102.92\\
\bottomrule
\end{tabular}
\end{table}
\endgroup

$n_0$: total of individuals with zero balance in HSA; p(k): k-th sample percentile; SD: standard deviation.

\medskip

\textbf{Table 5.} Descriptive statistics of the simulated individual mean HSA balance at every five years between 25 and 65 years-old. We present the mean ($\mu$) and standard deviation ($\sigma$) over the 1,000 simulations.

\begin{multicols}{2}  

According to the simulation, at 30 years-old 26\% of the lives had a zero balance in their HSA and, at 40 years-old, this percentage was lesser than 3\%, in average. As the annual health expenses are smaller at young age, the mean increase in the individual HSA balance is greater in early ages, and in later age ranges this balance is more spread around the mean. At 40 years-old, only 5\% of the lives had a balance greater than R\$ 30,000 in average and, at 45 years-old, between 50\% and 75\% of the lives had this kind of balance. At 55 years-old only 15\% of the lives had a balance lesser than R\$ 20,000, while  15\% had a balance greater than R\$ 50,000 in average and, at 60 years-old, only 25\% of the lives had a balance lesser than R\$ 30,000 in average and 50\% had a balance greater than R\$39,000 in average.

\begin{figure}[H]
    \includegraphics[scale=0.45]{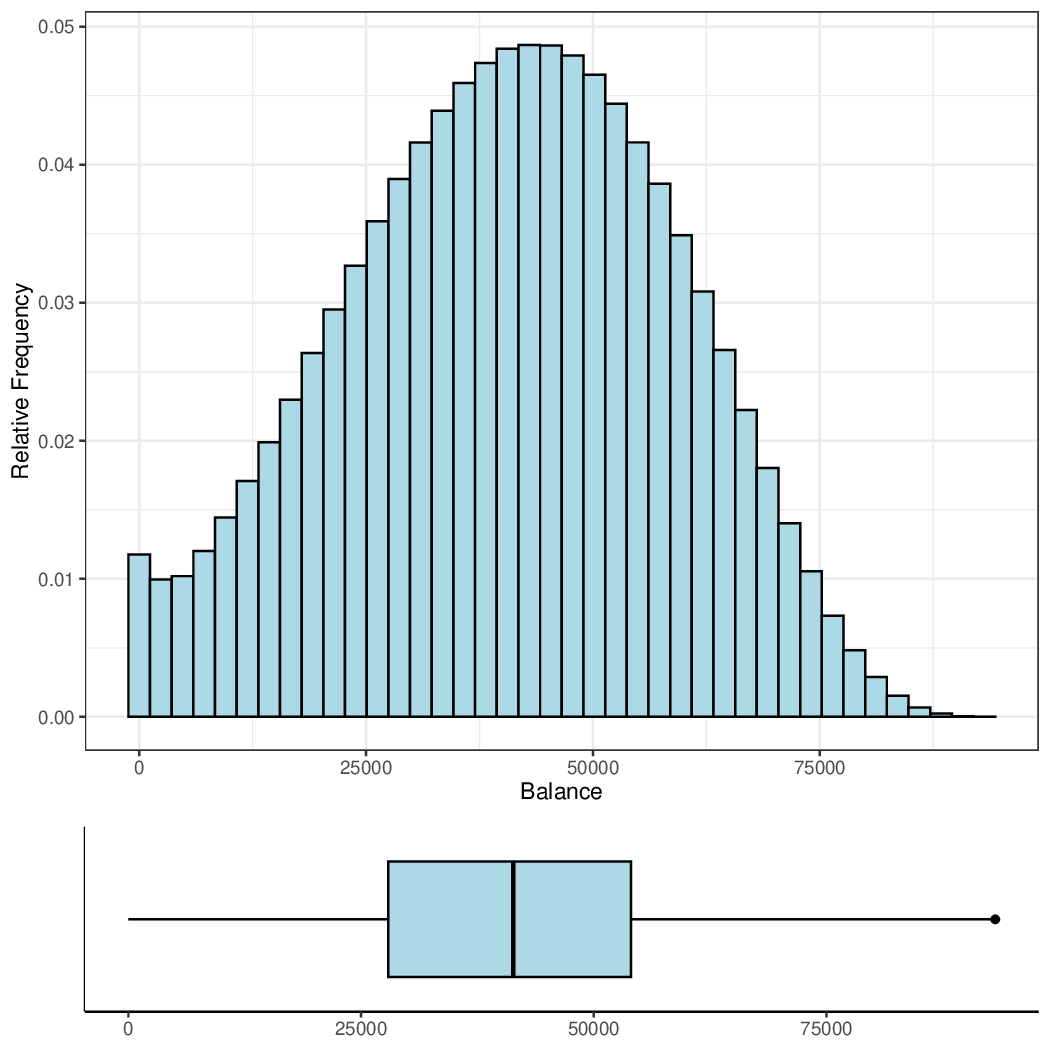}
    \caption{Empirical distribution of the simulated HSA balances at 65 years-old of the 10,000,000 simulated individuals.}
    \label{fig:empdist_1000sim}  % 
\end{figure}

At 65 years-old, the end of the work life, only 1\% of the lives had zero balance and 5\% a balance lesser than R\$11,000 in average, while half the lives had a balance greater than R\$41,000 in average, with 25\% with a balance greater than R\$53,000 in average. The mean balance at 65 years-old is around R\$40,000 in average, with a maximum balance around R\$89,000 in average. In Fig. \ref{fig:empdist_1000sim} we present the empirical distribution of the individual HSA balance in the 10,000,000 simulated individuals at 65 years-old. This distribution is approximately symmetrical (Skewness coefficient = -0.1202) and 01 outlier was detected (R\$93,142.83). According to the simulation, the HSA is a good alternative to fund healthcare costs at old age, as a great parcel of the simulated population had a considerable balance at retirement. Note that the simulation is based on the expenses of a portfolio in which there is no individual HSA, and that we considered a savings account without interest. Therefore, the balance of the individual HSA could in reality be greater than that simulated.

\subsection{Frequency and severity of the catastrophic insurance use}

In this section we analyze the use of the catastrophic insurance, more specifically its frequency and severity. In Table 6 we present the percentage (frequency), and the total value covered (severity), of the lives which used the catastrophic insurance each number of times.

\medskip

\end{multicols}

% latex table generated in R 4 2.3 by append_kable_to_file function
% Current Date/Time: 2024-02-25 01:05:40
\begingroup
\begin{table}[H]
\centering
\fontsize{7.5}{9.5}\selectfont
\begin{tabular}[t]{>{\centering\arraybackslash}p{0.8cm}>{\centering\arraybackslash}p{0.8cm}>{\centering\arraybackslash}p{0.8cm}>{\centering\arraybackslash}p{0.8cm}>{\centering\arraybackslash}p{0.8cm}>{\raggedleft\arraybackslash}p{1.6cm}>{\raggedleft\arraybackslash}p{1.6cm}>{\centering\arraybackslash}p{0.8cm}>{\centering\arraybackslash}p{0.8cm}>{\centering\arraybackslash}p{0.8cm}>{\raggedleft\arraybackslash}p{1.2cm}>{\raggedleft\arraybackslash}p{1.2cm}}
\toprule
\multicolumn{1}{c}{ } & \multicolumn{2}{c}{n} & \multicolumn{2}{c}{(\%) of Usage} & \multicolumn{2}{c}{Tot. CI Usage} & \multicolumn{3}{c}{ } & \multicolumn{2}{c}{Mean Usage per live} \\
\cmidrule(l{3pt}r{3pt}){2-3} \cmidrule(l{3pt}r{3pt}){4-5} \cmidrule(l{3pt}r{3pt}){6-7} \cmidrule(l{3pt}r{3pt}){11-12}
No. of
Times & $\mu$ & $\sigma$ & $\mu$ & $\sigma$ & $\mu$ & $\sigma$ & (\%) CI
Usage & Cum.(\%) of
Usage & Cum.(\%) CI
Usage & $\mu$ & $\sigma$\\
\midrule
\cellcolor{gray!10}{0} & \cellcolor{gray!10}{559} & \cellcolor{gray!10}{23.19} & \cellcolor{gray!10}{5.59} & \cellcolor{gray!10}{0.23} & \cellcolor{gray!10}{0} & \cellcolor{gray!10}{0} & \cellcolor{gray!10}{0.00} & \cellcolor{gray!10}{5.58} & \cellcolor{gray!10}{0.00} & \cellcolor{gray!10}{0} & \cellcolor{gray!10}{0}\\
1 & 1152 & 31.78 & 11.52 & 0.32 & 14,706,212 & 1,106,298 & 3.01 & 17.10 & 2.99 & 12,760 & 882\\
\cellcolor{gray!10}{2} & \cellcolor{gray!10}{1474} & \cellcolor{gray!10}{36.82} & \cellcolor{gray!10}{14.74} & \cellcolor{gray!10}{0.37} & \cellcolor{gray!10}{36,720,303} & \cellcolor{gray!10}{1,876,230} & \cellcolor{gray!10}{7.50} & \cellcolor{gray!10}{31.82} & \cellcolor{gray!10}{10.46} & \cellcolor{gray!10}{24,919} & \cellcolor{gray!10}{1,079}\\
3 & 1510 & 36.43 & 15.10 & 0.36 & 55,345,252 & 2,439,868 & 11.31 & 46.91 & 21.72 & 36,648 & 1,323\\
\cellcolor{gray!10}{4} & \cellcolor{gray!10}{1361} & \cellcolor{gray!10}{33.12} & \cellcolor{gray!10}{13.61} & \cellcolor{gray!10}{0.33} & \cellcolor{gray!10}{65,156,889} & \cellcolor{gray!10}{2,701,662} & \cellcolor{gray!10}{13.32} & \cellcolor{gray!10}{60.50} & \cellcolor{gray!10}{34.97} & \cellcolor{gray!10}{47,885} & \cellcolor{gray!10}{1,603}\\
\addlinespace
5 & 1121 & 32.67 & 11.21 & 0.33 & 65,797,506 & 2,987,801 & 13.45 & 71.71 & 48.35 & 58,667 & 1,975\\
\cellcolor{gray!10}{6} & \cellcolor{gray!10}{867} & \cellcolor{gray!10}{28.40} & \cellcolor{gray!10}{8.67} & \cellcolor{gray!10}{0.28} & \cellcolor{gray!10}{60,042,438} & \cellcolor{gray!10}{2,939,841} & \cellcolor{gray!10}{12.27} & \cellcolor{gray!10}{80.36} & \cellcolor{gray!10}{60.56} & \cellcolor{gray!10}{69,293} & \cellcolor{gray!10}{2,533}\\
7 & 636 & 24.31 & 6.36 & 0.24 & 50,631,601 & 2,721,570 & 10.35 & 86.72 & 70.86 & 79,608 & 3,089\\
\cellcolor{gray!10}{8} & \cellcolor{gray!10}{446} & \cellcolor{gray!10}{20.18} & \cellcolor{gray!10}{4.46} & \cellcolor{gray!10}{0.20} & \cellcolor{gray!10}{39,977,371} & \cellcolor{gray!10}{2,576,693} & \cellcolor{gray!10}{8.17} & \cellcolor{gray!10}{91.17} & \cellcolor{gray!10}{78.99} & \cellcolor{gray!10}{89,615} & \cellcolor{gray!10}{4,112}\\
9 & 302 & 16.89 & 3.02 & 0.17 & 29,969,236 & 2,245,427 & 6.12 & 94.19 & 85.09 & 99,320 & 5,125\\
\addlinespace
\cellcolor{gray!10}{10} & \cellcolor{gray!10}{200} & \cellcolor{gray!10}{14.02} & \cellcolor{gray!10}{2.00} & \cellcolor{gray!10}{0.14} & \cellcolor{gray!10}{21,602,668} & \cellcolor{gray!10}{2,021,014} & \cellcolor{gray!10}{4.41} & \cellcolor{gray!10}{96.19} & \cellcolor{gray!10}{89.48} & \cellcolor{gray!10}{108,085} & \cellcolor{gray!10}{6,792}\\
11 & 129 & 11.41 & 1.29 & 0.11 & 15,237,936 & 1,766,926 & 3.11 & 97.48 & 92.58 & 117,705 & 8,513\\
\cellcolor{gray!10}{12} & \cellcolor{gray!10}{84} & \cellcolor{gray!10}{9.04} & \cellcolor{gray!10}{0.84} & \cellcolor{gray!10}{0.09} & \cellcolor{gray!10}{10,595,814} & \cellcolor{gray!10}{1,449,420} & \cellcolor{gray!10}{2.16} & \cellcolor{gray!10}{98.32} & \cellcolor{gray!10}{94.73} & \cellcolor{gray!10}{125,683} & \cellcolor{gray!10}{10,782}\\
13 & 55 & 7.57 & 0.55 & 0.08 & 7,431,749 & 1,289,946 & 1.52 & 98.87 & 96.24 & 134,133 & 13,990\\
\cellcolor{gray!10}{14} & \cellcolor{gray!10}{37} & \cellcolor{gray!10}{6.02} & \cellcolor{gray!10}{0.37} & \cellcolor{gray!10}{0.06} & \cellcolor{gray!10}{5,205,901} & \cellcolor{gray!10}{1,061,262} & \cellcolor{gray!10}{1.06} & \cellcolor{gray!10}{99.24} & \cellcolor{gray!10}{97.30} & \cellcolor{gray!10}{141,704} & \cellcolor{gray!10}{17,071}\\
\addlinespace
15 & 24 & 4.78 & 0.24 & 0.05 & 3,651,611 & 915,835 & 0.75 & 99.48 & 98.05 & 149,867 & 21,897\\
\cellcolor{gray!10}{16} & \cellcolor{gray!10}{16} & \cellcolor{gray!10}{3.97} & \cellcolor{gray!10}{0.16} & \cellcolor{gray!10}{0.04} & \cellcolor{gray!10}{2,549,838} & \cellcolor{gray!10}{749,741} & \cellcolor{gray!10}{0.52} & \cellcolor{gray!10}{99.64} & \cellcolor{gray!10}{98.56} & \cellcolor{gray!10}{159,669} & \cellcolor{gray!10}{29,516}\\
17 & 10 & 3.32 & 0.10 & 0.03 & 1,747,222 & 693,394 & 0.36 & 99.75 & 98.92 & 166,207 & 37,701\\
\cellcolor{gray!10}{18} & \cellcolor{gray!10}{7} & \cellcolor{gray!10}{2.61} & \cellcolor{gray!10}{0.07} & \cellcolor{gray!10}{0.03} & \cellcolor{gray!10}{1,168,382} & \cellcolor{gray!10}{553,658} & \cellcolor{gray!10}{0.24} & \cellcolor{gray!10}{99.82} & \cellcolor{gray!10}{99.16} & \cellcolor{gray!10}{177,096} & \cellcolor{gray!10}{49,625}\\
19 & 4 & 1.88 & 0.04 & 0.02 & 744,571 & 416,481 & 0.15 & 99.86 & 99.31 & 181,694 & 60,022\\
\addlinespace
\cellcolor{gray!10}{20} & \cellcolor{gray!10}{3} & \cellcolor{gray!10}{1.43} & \cellcolor{gray!10}{0.03} & \cellcolor{gray!10}{0.01} & \cellcolor{gray!10}{504,427} & \cellcolor{gray!10}{338,090} & \cellcolor{gray!10}{0.10} & \cellcolor{gray!10}{99.88} & \cellcolor{gray!10}{99.41} & \cellcolor{gray!10}{192,019} & \cellcolor{gray!10}{91,845}\\
21 & 2 & 0.95 & 0.02 & 0.01 & 368,242 & 239,624 & 0.08 & 99.90 & 99.49 & 202,041 & 96,327\\
\cellcolor{gray!10}{22} & \cellcolor{gray!10}{1} & \cellcolor{gray!10}{0.67} & \cellcolor{gray!10}{0.01} & \cellcolor{gray!10}{0.01} & \cellcolor{gray!10}{308,767} & \cellcolor{gray!10}{210,895} & \cellcolor{gray!10}{0.06} & \cellcolor{gray!10}{99.91} & \cellcolor{gray!10}{99.55} & \cellcolor{gray!10}{219,688} & \cellcolor{gray!10}{128,687}\\
23 & 1 & 0.59 & 0.01 & 0.01 & 267,208 & 177,913 & 0.05 & 99.93 & 99.60 & 212,600 & 122,867\\
\cellcolor{gray!10}{24} & \cellcolor{gray!10}{1} & \cellcolor{gray!10}{0.32} & \cellcolor{gray!10}{0.01} & \cellcolor{gray!10}{0.00} & \cellcolor{gray!10}{260,201} & \cellcolor{gray!10}{173,980} & \cellcolor{gray!10}{0.05} & \cellcolor{gray!10}{99.94} & \cellcolor{gray!10}{99.66} & \cellcolor{gray!10}{231,627} & \cellcolor{gray!10}{137,556}\\
\addlinespace
25 & 1 & 0.25 & 0.01 & 0.00 & 257,426 & 140,819 & 0.05 & 99.95 & 99.71 & 241,421 & 119,099\\
\cellcolor{gray!10}{26} & \cellcolor{gray!10}{1} & \cellcolor{gray!10}{0.12} & \cellcolor{gray!10}{0.01} & \cellcolor{gray!10}{0.00} & \cellcolor{gray!10}{303,843} & \cellcolor{gray!10}{168,739} & \cellcolor{gray!10}{0.06} & \cellcolor{gray!10}{99.96} & \cellcolor{gray!10}{99.77} & \cellcolor{gray!10}{301,320} & \cellcolor{gray!10}{169,538}\\
27 & 1 & 0.21 & 0.01 & 0.00 & 349,976 & 283,372 & 0.07 & 99.97 & 99.84 & 339,873 & 283,236\\
\cellcolor{gray!10}{28} & \cellcolor{gray!10}{1} & \cellcolor{gray!10}{0.00} & \cellcolor{gray!10}{0.01} & \cellcolor{gray!10}{0.00} & \cellcolor{gray!10}{307,167} & \cellcolor{gray!10}{167,335} & \cellcolor{gray!10}{0.06} & \cellcolor{gray!10}{99.98} & \cellcolor{gray!10}{99.90} & \cellcolor{gray!10}{307,167} & \cellcolor{gray!10}{167,335}\\
29 & 1 & 0.00 & 0.01 & 0.00 & 237,283 & 109,358 & 0.05 & 99.99 & 99.95 & 237,283 & 109,358\\
\addlinespace
\cellcolor{gray!10}{30} & \cellcolor{gray!10}{1} & \cellcolor{gray!10}{0.00} & \cellcolor{gray!10}{0.01} & \cellcolor{gray!10}{0.00} & \cellcolor{gray!10}{239,257} & \cellcolor{gray!10}{71,911} & \cellcolor{gray!10}{0.05} & \cellcolor{gray!10}{100.00} & \cellcolor{gray!10}{100.00} & \cellcolor{gray!10}{239,257} & \cellcolor{gray!10}{71,911}\\
\bottomrule
\end{tabular}
\end{table}
\endgroup

\textbf{Table 6.} Mean Frequency and severity of the catastrophic insurance use by the simulated lives during the 40 years period. We present the mean ($\mu$) and standard deviation ($\sigma$) over the 1,000 simulations.

\begin{multicols}{2}  

In average, only 5.6\% of the simulated lives did not use the insurance at the 40 years period, approximately 50\% used at most three times, and 80\% used at most six times. For 0.5\% of the lives, it was necessary to use the insurance 13 times or more, and for 13 lives the insurance was triggered between 20 and 30 times among the 40 possibles. Furthermore, we note that around 50\% of the expenses covered by the catastrophic insurance were with lives that used it at most five times, which corresponds to 72\% of the lives, while 10\% of the expenses covered were with lives that used it 10 times or more, which corresponds to 3.5\% of lives in average. Also, around 0.5\% of the expenses covered by the insurance were with only  11 lives in average , which used it at least 21 times.

From a population point of view, the simulation of 1,000 repetitions of the portfolio during 40-years each, revealed a mean healthcare expense of R\$1,082,350,725 with standard deviation of R\$7,688,268, from which R\$489,367,594 (45.2\%) with standard deviation of R\$6,862,835 was covered by the catastrophic insurance (in average), and the remaining R\$592,983,131 (54.8\%) with standard deviation of R\$1,784,781 was covered by the individual HSA (in average).

\begin{figure}[H]
    \includegraphics[scale=0.45]{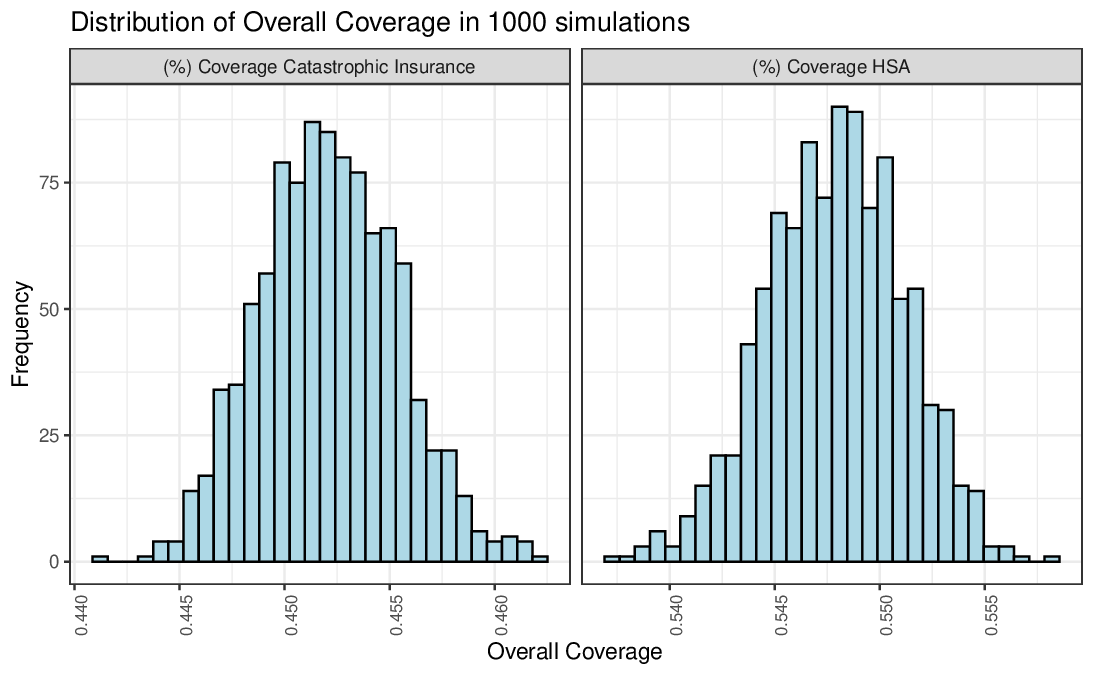}
    \caption{Distribution of percentage of coverage for healthcare expenses from HSA and catastrophic insurance in the 10,000,000 simulated individuals.}
    \label{fig:DistCov}  % 
\end{figure}

\medskip

In figure \ref{fig:DistCov} we present the distributions of percentage of coverage for healthcare expenses, either by HSA or catastrophic insurance.

\subsection{Savings account balance, coverage by HSA and coverage by catastrophic insurance}

The proposed CDHP product has three main features: the HSA balance at 65 years-old, the expenses covered by the HSA and the expenses covered by catastrophic insurance. The first two features are more susceptible to the effects of a HSA on an individual behavior, as the decision to consume healthcare goods and services may be inhibited by this CDHP product. Table 7 displays descriptive statistics of these three features.

\medskip

\end{multicols}

% latex table generated in R 4 2.3 by append_kable_to_file function
% Current Date/Time: 2024-02-29 14:12:27
\begingroup
\renewcommand{\thetable}{7}
\begin{table}[H]
\centering\begingroup\fontsize{10}{12}\selectfont

\begin{tabular}{>{\raggedright\arraybackslash}p{2cm}>{\raggedleft\arraybackslash}p{1.5cm}>{\raggedleft\arraybackslash}p{1.5cm}>{\raggedleft\arraybackslash}p{1.5cm}>{\raggedleft\arraybackslash}p{1.5cm}>{\raggedleft\arraybackslash}p{1.5cm}>{\raggedleft\arraybackslash}p{1.5cm}}
\toprule
\multicolumn{1}{c}{ } & \multicolumn{2}{c}{HSA Balance} & \multicolumn{2}{c}{HSA Coverage} & \multicolumn{2}{c}{Cat, Insurance Coverage} \\
\cmidrule(l{3pt}r{3pt}){2-3} \cmidrule(l{3pt}r{3pt}){4-5} \cmidrule(l{3pt}r{3pt}){6-7}
  & $\mu$ & $\sigma$ & $\mu$ & $\sigma$ & $\mu$ & $\sigma$\\
\midrule
\textbf{\cellcolor{gray!10}{pctnul}} & \cellcolor{gray!10}{0.73} & \cellcolor{gray!10}{-} & \cellcolor{gray!10}{0.00} & \cellcolor{gray!10}{-} & \cellcolor{gray!10}{5.59} & \cellcolor{gray!10}{-}\\
\textbf{p5} & 10,536 & 329 & 30,669 & 290 & 2,022 & 99\\
\textbf{\cellcolor{gray!10}{p10}} & \cellcolor{gray!10}{16,737} & \cellcolor{gray!10}{300} & \cellcolor{gray!10}{35,956} & \cellcolor{gray!10}{263} & \cellcolor{gray!10}{4,281} & \cellcolor{gray!10}{145}\\
\textbf{p15} & 21,267 & 288 & 39,834 & 259 & 6,638 & 177\\
\textbf{\cellcolor{gray!10}{p25}} & \cellcolor{gray!10}{28,263} & \cellcolor{gray!10}{271} & \cellcolor{gray!10}{45,996} & \cellcolor{gray!10}{249} & \cellcolor{gray!10}{11,970} & \cellcolor{gray!10}{249}\\
\addlinespace
\textbf{p40} & 36,568 & 256 & 53,738 & 241 & 21,650 & 363\\
\textbf{\cellcolor{gray!10}{p50}} & \cellcolor{gray!10}{41,535} & \cellcolor{gray!10}{248} & \cellcolor{gray!10}{58,645} & \cellcolor{gray!10}{248} & \cellcolor{gray!10}{29,666} & \cellcolor{gray!10}{459}\\
\textbf{p75} & 54,106 & 249 & 72,079 & 274 & 63,926 & 948\\
\textbf{\cellcolor{gray!10}{p85}} & \cellcolor{gray!10}{60,242} & \cellcolor{gray!10}{257} & \cellcolor{gray!10}{79,243} & \cellcolor{gray!10}{298} & \cellcolor{gray!10}{94,228} & \cellcolor{gray!10}{1,586}\\
\textbf{p95} & 69,381 & 290 & 90,573 & 357 & 178,278 & 4,494\\
\addlinespace
\textbf{\cellcolor{gray!10}{p98}} & \cellcolor{gray!10}{74,616} & \cellcolor{gray!10}{342} & \cellcolor{gray!10}{96,831} & \cellcolor{gray!10}{359} & \cellcolor{gray!10}{264,692} & \cellcolor{gray!10}{6,711}\\
\textbf{max} & 89,251 & 1,355 & 100,000 & - & 1,176,242 & 119,167\\
\textbf{\cellcolor{gray!10}{mean}} & \cellcolor{gray!10}{41,000} & \cellcolor{gray!10}{176} & \cellcolor{gray!10}{59,298} & \cellcolor{gray!10}{178} & \cellcolor{gray!10}{51,833} & \cellcolor{gray!10}{713}\\
\textbf{sd} & 17,739 & 108 & 18,015 & 111 & 69,719 & 2,297\\
\bottomrule
\end{tabular}
\endgroup{}
\end{table}
\endgroup

p(k): k-th sample percentile; SD: standard deviation.

\medskip

\textbf{Table 7.} Descriptive statistics of the HSA balance at 65 years-old, the expenses covered by the HSA and the expenses covered by catastrophic insurance over the work life. The percentiles, mean and standard deviation are calculated considering only the values greater than zero (n=10,000). We also present the mean ($\mu$) and standard deviation ($\sigma$) over the 1,000 simulations.

\begin{multicols}{2}  

If a life did not have any expenses in the 40 years period, its account balance would be R\$ 100,000, although the simulated mean and median balance at 65 years-old were between R\$41,000 and R\$41,535, in average with s.d. of R\$248, and the maximum was R\$ 89,251 in average, with s.d. of R\$ 1,355. Therefore, the mean account balance at retirement is in average 42\% the value which has been deposited. The mean expense covered during the work life by the insurance was around R\$51,800, in average with s.d. of R\$713, and the maximum was around R\$1,176,000, in average with s.d. of R\$ 119,167, which is 22 times the mean value, evidencing the presence of outliers. For 25\% of the lives, the insurance covered lesser than R\$ 12,000, and for half the lives it covered at most R\$29,700 in average with s.d. of R\$459. On the other hand, the 5\% of lives with the greatest expenses covered by the insurance had more than R\$ 175,000 covered in average. When considering the percentage of expenses covered by the insurance, half the lives had at most 40\% of them covered by the insurance in average, while 25\% of the lives had almost half of their expenses covered by it in average, and 5\% of the lives had more than 75\% of their expenses covered by the insurance in average.

\begin{figure}[H]
    \includegraphics[scale=0.45]{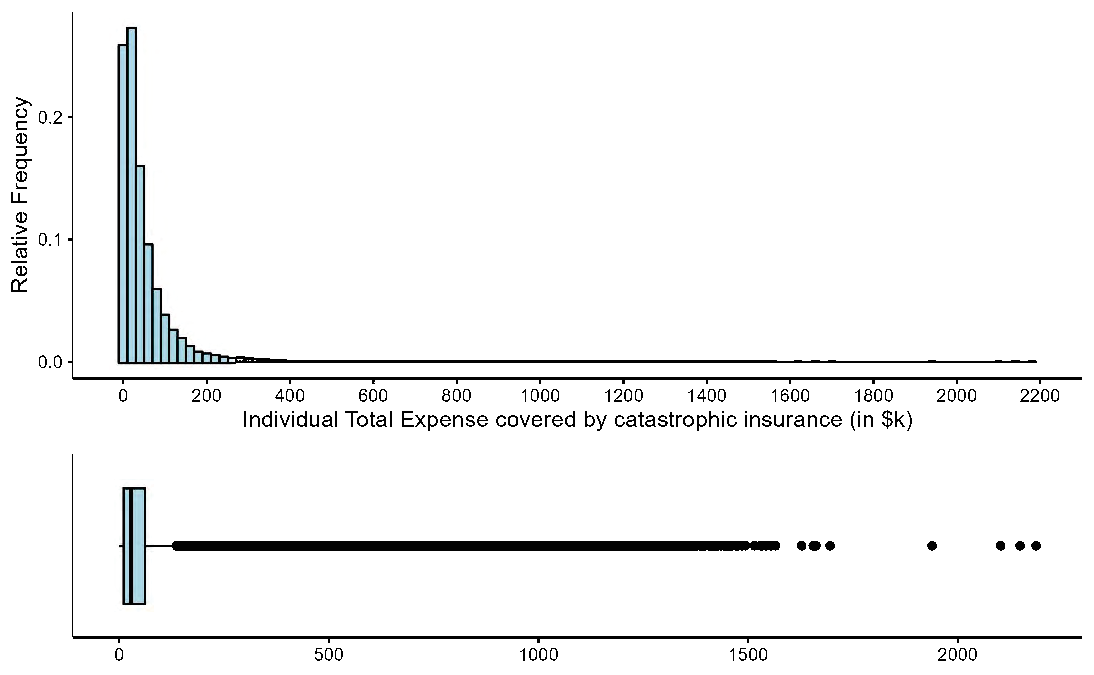}
    \caption{Severity of catastrophic insurance use in the 10,000,000 simulated individuals.}
    \label{fig:IndExp_CI}  % 
\end{figure}

\begin{figure}[H]
    \includegraphics[scale=0.45]{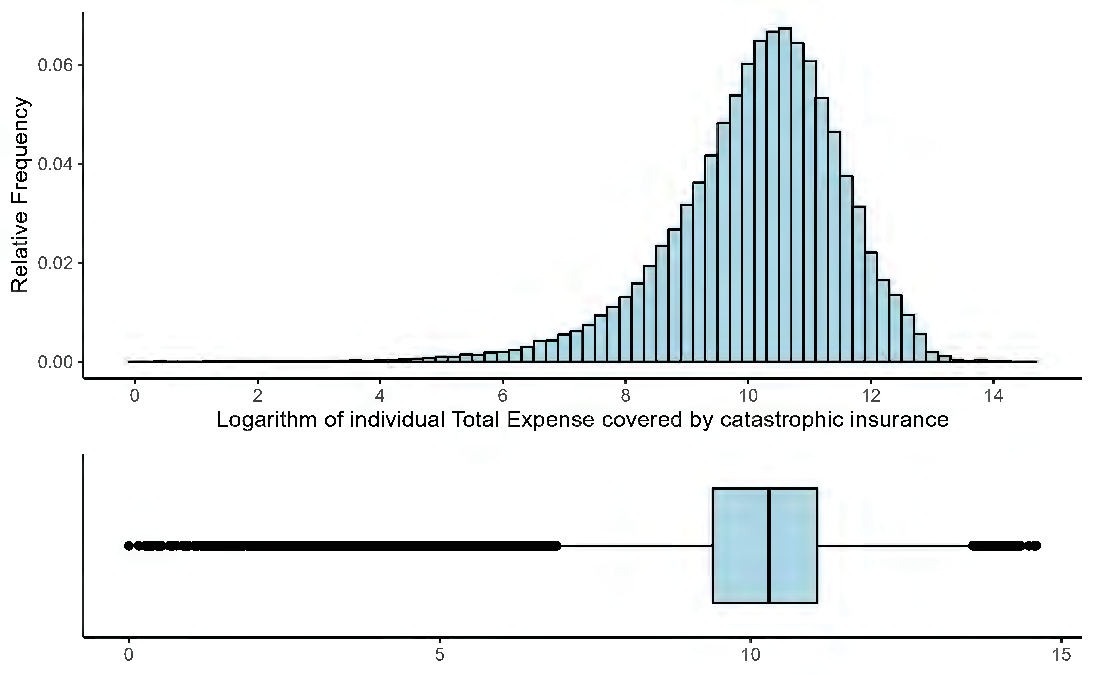}
    \caption{Logarithm transform of severity of catastrophic insurance usage in the 10,000,000 simulated individuals.}
    \label{fig:LogIndExp_CI}  % 
\end{figure}

\medskip

Figures \ref{fig:IndExp_CI} and \ref{fig:LogIndExp_CI} display the empirical distribution of the severity of catastrophic insurance use, and of its logarithm transformation, where it can be noted the presence of outliers, evidencing the coverage of great healthcare expenses by the insurance. Coverages under R\$1.00 were ignored in the logarithmic transformation.

\begin{figure}[H]
    \includegraphics[scale=0.38]{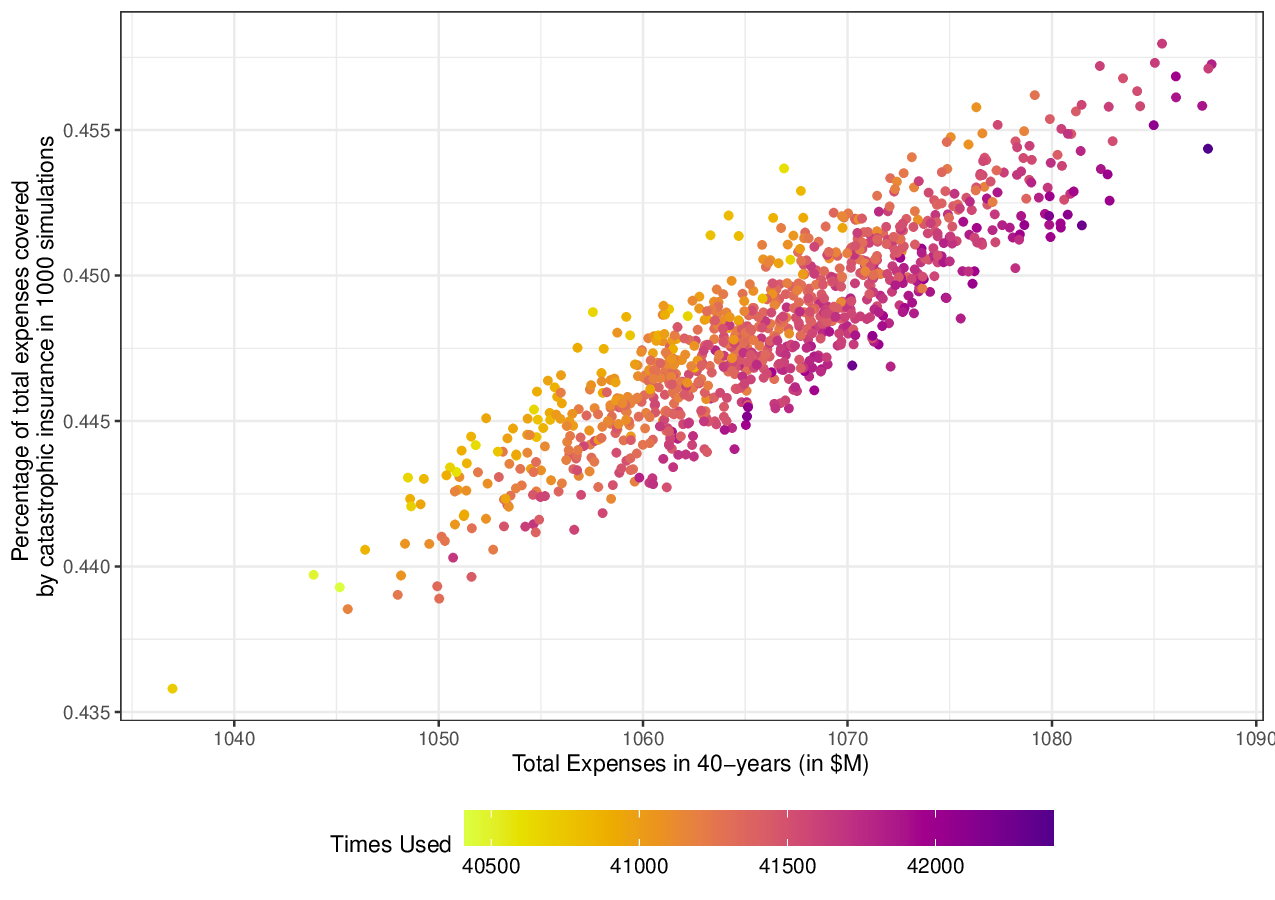}
    \caption{Dispersion plot between the total expenses at the 40 year period of each life in  the 10,000,000 simulated individuals and the percentage of these expenses which were covered by catastrophic insurance. The color refers to the number of times in which the insurance was used over the work life.}
    \label{fig:PctExpCov_CI}  % 
\end{figure}

\medskip

In Figure \ref{fig:PctExpCov_CI} we present the dispersion plot between the total expenses at the 40 year period of each life and the percentage of these expenses which were covered by catastrophic insurance. The color of the points refers to the number of times in which the insurance was used over the work life.

\section{Discussion and Conclusions}

An increase in healthcare costs has been observed recently in several countries, especially in Brazil, where it has led to a discussion about co-payment and deductible devices for health plans, which culminated in a resolution by the regulatory agency (ANS) to incorporate these devices to the Brazilian supplementary health system. A negative public repercussion based on misunderstandings followed the resolution, forcing ANS to revoke it, leaving the market uncovered regarding the understanding of these devices. In this context, this paper aimed to subsidize the discussion about co-payment and deductibles by simulating a consumer directed health plan product which has never been commercialized in Brazil, the health savings accounts, which combines a high deductible health plan with tax breaks to form a savings account aiming to reduce the individual exposition to the risk of extremal events by combining a catastrophic insurance with an individual savings account.

This mixed product benefits both the employer and employee. From the employee point of view, the accredited network of healthcare providers is superior than the standard, and the individual is not limited to the accredited network. Moreover, the balance may be invested to increase income, but due to the purpose of this work, this issue of investment has not been considered. From the employer point of view, it is expected lower individual healthcare expenses, since the employees, when paying for their own healthcare goods and services, tend to be more selective in choosing their providers and more thrifty when using the available services.

In order to carry out our simulation, we had to first establish a prediction technique for annual healthcare expenses. As evidenced in the literature, this is not an easy endeavor due to peculiarities of healthcare expenses data: great number of correlated independent variables, enormous variability, non-linearity, excess of zeros and presence of extreme values \cite{Duncan_etal2016}. In order to circumvent these difficulties, we proposed an approach to healthcare expenses prediction based on Markov Chains, which seeks to predict the annual expenses of an individual based on his or her sex, age range and expenses in the last two years. In this approach, first the expense level is estimated and then, the exact expense is chosen from an empirical distribution.

Although classical in Statistics and Probability Theory, and applied in health economics to predict health conditions and costs related to it (\cite{Komorowski_Raffa2016}, \cite{Sato_Zouain2010}, \cite{Jeong_etal2014}, \cite{Graves_etal2016}, \cite{Bala_Mauskopf2006}, \cite{Dobi_Zempleni2019} and \cite{Garg_etal2010}), it seems that a Markov Chain approach may also be useful to predict annual healthcare expenses, as it is known that expenses history is an important factor in predicting healthcare expenses \cite{Duncan_etal2016}. A Markov Chain approach based on previous expenses, sex, and age range may be more viable for pricing health plans than a method based on health condition due to ethical issues.

Based on a dataset containing annual expenses of a portfolio of a Brazilian self-management health plan, we simulated the proposed HSA product for 10,000 lives from 25 to 65 years-old in order to study the balance of the individual savings accounts over time and at retirement, and the frequency and severity of catastrophic insurance use. In this simulation, we took into account the persistence of costs phenomenon, widely known to be part of individual healthcare expenses. The results evidenced a low prevalence of persistence of costs at the individual level, as at 65 years-old the balance of the individual savings accounts is symmetrically spread around the average of R\$ 44,000, supporting that the adoption of this model is not a mechanism of inequalities generation. During the 40 years period in the 10,000,000 simulated individuals, the total healthcare expenses of the simulated portfolio was R\$1,082,350,725 in average with s.d. of R\$7,688,268, of which R\$489,367,594 in average (45.2\%) with s.d. R\$6,862,835 was covered by catastrophic insurance. From the individual perspective, half the lives had 40\% of their expenses covered by catastrophic insurance and only 5\% of them had more than 75\% of their expenses covered by catastrophic insurance. At retirement, the mean balance of the individual accounts was 45.2\% of the value deposited over the work life, so the mean expense covered by the HSA during the work life was 54.8\% that deposited, in average.

Therefore, we may conclude that a mixed CDHP product combining a HSA and a catastrophic insurance may be viable in Brazil from a population point of view as it does not create inequalities caused by persistence of costs. In order to implement such a product, one needs to adapt the value deposited by the employer every year and the limit covered by the HSA to suit the health system costs. 

We leave some topics for future research. An interesting topic would be to compare a Markov Chain approach to healthcare expenses prediction with the usual methods based on regression. Also, the development of a method which incorporates the best features of both approaches seems promising. About the proposed CDHP, there are some hyperparameters which can be optimized in order to obtain a more realistic simulation: the break points of the expense levels, the value deposited by the employer every year and the limit of expenses covered by the HSA. In order to optimize these quantities, it would be needed data with better quality and greater quantity than that of this paper. Finally, to further the discussion in Brazil about co-payment and deductible devices, it would be interesting to study other kinds of CDHP products in the Brazilian context.

\end{multicols} 

\clearpage

\pagebreak

\onecolumn

\section*{Appendix A - Descriptive Statistics and Transition Matrix}

\subsection*{Descriptive Statistics of Annual Expenses, by sex}

\medskip

In this section, it is presented descriptive statistics by sex of the annual expenses individuals by sex, between 25 and 65 years-old which stayed on the portfolio during the whole five years period. The percentiles, mean and standard deviation are calculated considering only the individuals with positive expenses.

% latex table generated in R 4 2.3 by append_kable_to_file function
% Current Date/Time: 2024-02-18 00:57:56
\begingroup
\renewcommand{\thetable}{2.1}
\begin{table}[H]
\centering
\caption{Descriptive Statistics Annual Expenses (Female) }
\centering
\begin{tabular}[t]{>{}l>{\raggedleft\arraybackslash}p{2.5cm}>{\raggedleft\arraybackslash}p{2.5cm}>{\raggedleft\arraybackslash}p{2.5cm}>{\raggedleft\arraybackslash}p{2.5cm}>{\raggedleft\arraybackslash}p{2.5cm}}
\toprule
Description & 2005 & 2006 & 2007 & 2008 & 2009\\
\midrule
\textbf{\cellcolor{gray!10}{n}} & \cellcolor{gray!10}{13,539} & \cellcolor{gray!10}{13,539} & \cellcolor{gray!10}{13,539} & \cellcolor{gray!10}{13,539} & \cellcolor{gray!10}{13,539}\\
\textbf{PctNoExpense} & 5.09 & 5.26 & 5.16 & 5.15 & 5.55\\
\textbf{\cellcolor{gray!10}{p25}} & \cellcolor{gray!10}{457} & \cellcolor{gray!10}{422} & \cellcolor{gray!10}{449} & \cellcolor{gray!10}{487} & \cellcolor{gray!10}{490}\\
\textbf{p50} & 998 & 940 & 987 & 1,130 & 1,104\\
\textbf{\cellcolor{gray!10}{p75}} & \cellcolor{gray!10}{2,193} & \cellcolor{gray!10}{2,029} & \cellcolor{gray!10}{2,152} & \cellcolor{gray!10}{2,539} & \cellcolor{gray!10}{2,464}\\
\addlinespace
\textbf{p90} & 4,808 & 4,462 & 4,848 & 5,603 & 5,453\\
\textbf{\cellcolor{gray!10}{p95}} & \cellcolor{gray!10}{8,402} & \cellcolor{gray!10}{7,501} & \cellcolor{gray!10}{8,247} & \cellcolor{gray!10}{9,338} & \cellcolor{gray!10}{9,662}\\
\textbf{p96} & 9,781 & 8,818 & 9,393 & 10,659 & 11,122\\
\textbf{\cellcolor{gray!10}{p97}} & \cellcolor{gray!10}{11,293} & \cellcolor{gray!10}{10,376} & \cellcolor{gray!10}{11,416} & \cellcolor{gray!10}{12,757} & \cellcolor{gray!10}{13,915}\\
\textbf{p98} & 14,425 & 13,252 & 15,886 & 17,114 & 19,095\\
\addlinespace
\textbf{\cellcolor{gray!10}{p99}} & \cellcolor{gray!10}{24,076} & \cellcolor{gray!10}{21,371} & \cellcolor{gray!10}{24,681} & \cellcolor{gray!10}{27,696} & \cellcolor{gray!10}{30,718}\\
\textbf{p995} & 33,027 & 34,656 & 44,429 & 51,048 & 53,649\\
\textbf{\cellcolor{gray!10}{p999}} & \cellcolor{gray!10}{68,154} & \cellcolor{gray!10}{125,250} & \cellcolor{gray!10}{98,976} & \cellcolor{gray!10}{137,450} & \cellcolor{gray!10}{203,682}\\
\textbf{max} & 282,993 & 528,293 & 355,220 & 1,044,525 & 550,309\\
\textbf{\cellcolor{gray!10}{mean}} & \cellcolor{gray!10}{2,341} & \cellcolor{gray!10}{2,327} & \cellcolor{gray!10}{2,494} & \cellcolor{gray!10}{2,956} & \cellcolor{gray!10}{3,062}\\
\addlinespace
\textbf{sd} & 6,068 & 8,897 & 7,828 & 13,247 & 11,958\\
\bottomrule
\end{tabular}
\label{tab:table2.1}
\end{table}
\endgroup
\textbf{n}: total of individuals in portfolio;\textbf{p(k)}: k-th sample percentile;\textbf{SD}: standard deviation

\medskip

% latex table generated in R 4 2.3 by append_kable_to_file function
% Current Date/Time: 2024-02-18 00:57:57
\begingroup
\renewcommand{\thetable}{2.2}
\begin{table}[H]
\centering
\caption{Descriptive Statistics Annual Expenses (Male) }
\centering
\begin{tabular}[t]{>{}l>{\raggedleft\arraybackslash}p{2.5cm}>{\raggedleft\arraybackslash}p{2.5cm}>{\raggedleft\arraybackslash}p{2.5cm}>{\raggedleft\arraybackslash}p{2.5cm}>{\raggedleft\arraybackslash}p{2.5cm}}
\toprule
Description & 2005 & 2006 & 2007 & 2008 & 2009\\
\midrule
\textbf{\cellcolor{gray!10}{n}} & \cellcolor{gray!10}{14,241} & \cellcolor{gray!10}{14,241} & \cellcolor{gray!10}{14,241} & \cellcolor{gray!10}{14,241} & \cellcolor{gray!10}{14,241}\\
\textbf{PctNoExpense} & 6.81 & 5.77 & 6.35 & 5.93 & 5.97\\
\textbf{\cellcolor{gray!10}{p25}} & \cellcolor{gray!10}{237} & \cellcolor{gray!10}{231} & \cellcolor{gray!10}{252} & \cellcolor{gray!10}{289} & \cellcolor{gray!10}{271}\\
\textbf{p50} & 580 & 572 & 625 & 721 & 681\\
\textbf{\cellcolor{gray!10}{p75}} & \cellcolor{gray!10}{1,351} & \cellcolor{gray!10}{1,336} & \cellcolor{gray!10}{1,432} & \cellcolor{gray!10}{1,677} & \cellcolor{gray!10}{1,594}\\
\addlinespace
\textbf{p90} & 2,983 & 2,914 & 3,208 & 3,927 & 3,868\\
\textbf{\cellcolor{gray!10}{p95}} & \cellcolor{gray!10}{5,588} & \cellcolor{gray!10}{5,491} & \cellcolor{gray!10}{6,259} & \cellcolor{gray!10}{7,680} & \cellcolor{gray!10}{8,212}\\
\textbf{p96} & 6,847 & 6,838 & 7,833 & 9,690 & 10,494\\
\textbf{\cellcolor{gray!10}{p97}} & \cellcolor{gray!10}{8,799} & \cellcolor{gray!10}{8,865} & \cellcolor{gray!10}{10,461} & \cellcolor{gray!10}{12,985} & \cellcolor{gray!10}{14,589}\\
\textbf{p98} & 12,720 & 12,836 & 14,960 & 19,653 & 21,593\\
\addlinespace
\textbf{\cellcolor{gray!10}{p99}} & \cellcolor{gray!10}{22,163} & \cellcolor{gray!10}{21,840} & \cellcolor{gray!10}{26,596} & \cellcolor{gray!10}{34,427} & \cellcolor{gray!10}{36,397}\\
\textbf{p995} & 35,525 & 36,021 & 42,975 & 59,041 & 57,155\\
\textbf{\cellcolor{gray!10}{p999}} & \cellcolor{gray!10}{100,317} & \cellcolor{gray!10}{100,298} & \cellcolor{gray!10}{90,679} & \cellcolor{gray!10}{139,391} & \cellcolor{gray!10}{146,580}\\
\textbf{max} & 426,772 & 371,231 & 266,119 & 286,716 & 965,287\\
\textbf{\cellcolor{gray!10}{mean}} & \cellcolor{gray!10}{1,795} & \cellcolor{gray!10}{1,768} & \cellcolor{gray!10}{1,956} & \cellcolor{gray!10}{2,473} & \cellcolor{gray!10}{2,576}\\
\addlinespace
\textbf{sd} & 7,490 & 7,306 & 7,235 & 9,786 & 13,316\\
\bottomrule
\end{tabular}
\label{tab:table2.2}
\end{table}
\endgroup
\textbf{n}: total of individuals in portfolio;\textbf{p(k)}: k-th sample percentile;\textbf{SD}: standard deviation

\subsection*{Descriptive Statistics of Annual Expenses, by sex and age category}

\medskip

In this section, it is presented descriptive statistics by age range of the annual expenses of individuals between 25 and 65 years-old which stayed on the portfolio during the whole five years period. The percentiles, mean and standard deviation are calculated considering only the individuals with positive expenses.

% latex table generated in R 4 2.3 by append_kable_to_file function
% Current Date/Time: 2024-02-18 01:18:01
\begingroup
\renewcommand{\thetable}{3.1}
\begin{table}[H]
\centering
\caption{Descriptive Statistics Annual Expenses (Female) 21a24}
\centering
\begin{tabular}[t]{>{}l>{\raggedleft\arraybackslash}p{2.5cm}>{\raggedleft\arraybackslash}p{2.5cm}>{\raggedleft\arraybackslash}p{2.5cm}>{\raggedleft\arraybackslash}p{2.5cm}}
\toprule
Description & 2005 & 2006 & 2007 & 2008\\
\midrule
\textbf{\cellcolor{gray!10}{n}} & \cellcolor{gray!10}{555} & \cellcolor{gray!10}{372} & \cellcolor{gray!10}{210} & \cellcolor{gray!10}{75}\\
\textbf{PctNoExpense} & 1.98 & 3.76 & 5.24 & 9.33\\
\textbf{\cellcolor{gray!10}{p25}} & \cellcolor{gray!10}{339} & \cellcolor{gray!10}{349} & \cellcolor{gray!10}{390} & \cellcolor{gray!10}{291}\\
\textbf{p50} & 833 & 730 & 888 & 699\\
\textbf{\cellcolor{gray!10}{p75}} & \cellcolor{gray!10}{1,974} & \cellcolor{gray!10}{1,740} & \cellcolor{gray!10}{2,278} & \cellcolor{gray!10}{1,334}\\
\addlinespace
\textbf{p90} & 4,661 & 3,678 & 5,621 & 3,088\\
\textbf{\cellcolor{gray!10}{p95}} & \cellcolor{gray!10}{6,495} & \cellcolor{gray!10}{5,171} & \cellcolor{gray!10}{9,039} & \cellcolor{gray!10}{6,091}\\
\textbf{p96} & 7,803 & 6,268 & 11,516 & 6,626\\
\textbf{\cellcolor{gray!10}{p97}} & \cellcolor{gray!10}{8,963} & \cellcolor{gray!10}{7,937} & \cellcolor{gray!10}{12,664} & \cellcolor{gray!10}{6,749}\\
\textbf{p98} & 10,295 & 10,176 & 19,717 & 8,433\\
\addlinespace
\textbf{\cellcolor{gray!10}{p99}} & \cellcolor{gray!10}{14,738} & \cellcolor{gray!10}{11,447} & \cellcolor{gray!10}{29,037} & \cellcolor{gray!10}{10,056}\\
\textbf{p995} & 19,713 & 15,185 & 30,401 & 10,823\\
\textbf{\cellcolor{gray!10}{p999}} & \cellcolor{gray!10}{74,328} & \cellcolor{gray!10}{32,586} & \cellcolor{gray!10}{63,712} & \cellcolor{gray!10}{11,437}\\
\textbf{max} & 130,804 & 41,884 & 72,040 & 11,591\\
\textbf{\cellcolor{gray!10}{mean}} & \cellcolor{gray!10}{2,030} & \cellcolor{gray!10}{1,617} & \cellcolor{gray!10}{2,609} & \cellcolor{gray!10}{1,410}\\
\addlinespace
\textbf{sd} & 6,198 & 3,051 & 6,446 & 2,122\\
\bottomrule
\end{tabular}
\label{tab:table3.1}
\end{table}
\endgroup
\textbf{n}: total of individuals in portfolio;\textbf{p(k)}: k-th sample percentile;\textbf{SD}: standard deviation

\medskip

% latex table generated in R 4 2.3 by append_kable_to_file function
% Current Date/Time: 2024-02-18 01:18:01
\begingroup
\renewcommand{\thetable}{3.2}
\begin{table}[H]
\centering
\caption{Descriptive Statistics Annual Expenses (Male) 21a24}
\centering
\begin{tabular}[t]{>{}l>{\raggedleft\arraybackslash}p{2.5cm}>{\raggedleft\arraybackslash}p{2.5cm}>{\raggedleft\arraybackslash}p{2.5cm}>{\raggedleft\arraybackslash}p{2.5cm}}
\toprule
Description & 2005 & 2006 & 2007 & 2008\\
\midrule
\textbf{\cellcolor{gray!10}{n}} & \cellcolor{gray!10}{375} & \cellcolor{gray!10}{250} & \cellcolor{gray!10}{149} & \cellcolor{gray!10}{65}\\
\textbf{PctNoExpense} & 10.93 & 12.00 & 14.77 & 15.38\\
\textbf{\cellcolor{gray!10}{p25}} & \cellcolor{gray!10}{182} & \cellcolor{gray!10}{127} & \cellcolor{gray!10}{139} & \cellcolor{gray!10}{123}\\
\textbf{p50} & 415 & 322 & 366 & 296\\
\textbf{\cellcolor{gray!10}{p75}} & \cellcolor{gray!10}{997} & \cellcolor{gray!10}{803} & \cellcolor{gray!10}{809} & \cellcolor{gray!10}{848}\\
\addlinespace
\textbf{p90} & 2,338 & 1,710 & 2,080 & 1,682\\
\textbf{\cellcolor{gray!10}{p95}} & \cellcolor{gray!10}{6,259} & \cellcolor{gray!10}{3,174} & \cellcolor{gray!10}{3,354} & \cellcolor{gray!10}{1,891}\\
\textbf{p96} & 8,641 & 3,473 & 4,105 & 1,938\\
\textbf{\cellcolor{gray!10}{p97}} & \cellcolor{gray!10}{10,881} & \cellcolor{gray!10}{3,739} & \cellcolor{gray!10}{4,756} & \cellcolor{gray!10}{3,092}\\
\textbf{p98} & 13,469 & 4,524 & 8,094 & 4,712\\
\addlinespace
\textbf{\cellcolor{gray!10}{p99}} & \cellcolor{gray!10}{16,587} & \cellcolor{gray!10}{7,974} & \cellcolor{gray!10}{10,062} & \cellcolor{gray!10}{10,423}\\
\textbf{p995} & 25,255 & 20,678 & 10,517 & 13,634\\
\textbf{\cellcolor{gray!10}{p999}} & \cellcolor{gray!10}{94,741} & \cellcolor{gray!10}{23,561} & \cellcolor{gray!10}{10,945} & \cellcolor{gray!10}{16,203}\\
\textbf{max} & 128,157 & 24,017 & 11,052 & 16,846\\
\textbf{\cellcolor{gray!10}{mean}} & \cellcolor{gray!10}{1,688} & \cellcolor{gray!10}{889} & \cellcolor{gray!10}{923} & \cellcolor{gray!10}{899}\\
\addlinespace
\textbf{sd} & 7,596 & 2,362 & 1,762 & 2,330\\
\bottomrule
\end{tabular}
\label{tab:table3.2}
\end{table}
\endgroup
\textbf{n}: total of individuals in portfolio;\textbf{p(k)}: k-th sample percentile;\textbf{SD}: standard deviation

\medskip

% latex table generated in R 4 2.3 by append_kable_to_file function
% Current Date/Time: 2024-02-18 01:18:01
\begingroup
\renewcommand{\thetable}{3.3}
\begin{table}[H]
\centering
\caption{Descriptive Statistics Annual Expenses (Female) 25a30}
\centering
\begin{tabular}[t]{>{}l>{\raggedleft\arraybackslash}p{2.5cm}>{\raggedleft\arraybackslash}p{2.5cm}>{\raggedleft\arraybackslash}p{2.5cm}>{\raggedleft\arraybackslash}p{2.5cm}>{\raggedleft\arraybackslash}p{2.5cm}}
\toprule
Description & 2005 & 2006 & 2007 & 2008 & 2009\\
\midrule
\textbf{\cellcolor{gray!10}{n}} & \cellcolor{gray!10}{1,541} & \cellcolor{gray!10}{1,371} & \cellcolor{gray!10}{1,267} & \cellcolor{gray!10}{1,160} & \cellcolor{gray!10}{1,002}\\
\textbf{PctNoExpense} & 4.67 & 4.60 & 3.95 & 4.31 & 5.29\\
\textbf{\cellcolor{gray!10}{p25}} & \cellcolor{gray!10}{393} & \cellcolor{gray!10}{328} & \cellcolor{gray!10}{330} & \cellcolor{gray!10}{372} & \cellcolor{gray!10}{358}\\
\textbf{p50} & 907 & 737 & 769 & 880 & 880\\
\textbf{\cellcolor{gray!10}{p75}} & \cellcolor{gray!10}{2,427} & \cellcolor{gray!10}{1,751} & \cellcolor{gray!10}{1,797} & \cellcolor{gray!10}{2,011} & \cellcolor{gray!10}{1,873}\\
\addlinespace
\textbf{p90} & 5,535 & 4,493 & 4,694 & 5,093 & 4,647\\
\textbf{\cellcolor{gray!10}{p95}} & \cellcolor{gray!10}{8,885} & \cellcolor{gray!10}{8,096} & \cellcolor{gray!10}{7,014} & \cellcolor{gray!10}{8,777} & \cellcolor{gray!10}{9,509}\\
\textbf{p96} & 9,876 & 8,901 & 7,996 & 10,444 & 10,602\\
\textbf{\cellcolor{gray!10}{p97}} & \cellcolor{gray!10}{10,801} & \cellcolor{gray!10}{10,246} & \cellcolor{gray!10}{8,702} & \cellcolor{gray!10}{11,730} & \cellcolor{gray!10}{12,047}\\
\textbf{p98} & 12,642 & 11,821 & 10,653 & 13,740 & 14,197\\
\addlinespace
\textbf{\cellcolor{gray!10}{p99}} & \cellcolor{gray!10}{17,993} & \cellcolor{gray!10}{17,090} & \cellcolor{gray!10}{14,164} & \cellcolor{gray!10}{18,273} & \cellcolor{gray!10}{19,864}\\
\textbf{p995} & 24,900 & 21,846 & 27,135 & 29,606 & 27,090\\
\textbf{\cellcolor{gray!10}{p999}} & \cellcolor{gray!10}{52,313} & \cellcolor{gray!10}{38,304} & \cellcolor{gray!10}{83,367} & \cellcolor{gray!10}{74,171} & \cellcolor{gray!10}{42,511}\\
\textbf{max} & 85,433 & 52,910 & 118,517 & 80,355 & 57,494\\
\textbf{\cellcolor{gray!10}{mean}} & \cellcolor{gray!10}{2,279} & \cellcolor{gray!10}{1,887} & \cellcolor{gray!10}{2,002} & \cellcolor{gray!10}{2,223} & \cellcolor{gray!10}{2,114}\\
\addlinespace
\textbf{sd} & 4,416 & 3,656 & 5,623 & 5,058 & 4,204\\
\bottomrule
\end{tabular}
\label{tab:table3.3}
\end{table}
\endgroup
\textbf{n}: total of individuals in portfolio;\textbf{p(k)}: k-th sample percentile;\textbf{SD}: standard deviation

\medskip

% latex table generated in R 4 2.3 by append_kable_to_file function
% Current Date/Time: 2024-02-18 01:18:01
\begingroup
\renewcommand{\thetable}{3.4}
\begin{table}[H]
\centering
\caption{Descriptive Statistics Annual Expenses (Male) 25a30}
\centering
\begin{tabular}[t]{>{}l>{\raggedleft\arraybackslash}p{2.5cm}>{\raggedleft\arraybackslash}p{2.5cm}>{\raggedleft\arraybackslash}p{2.5cm}>{\raggedleft\arraybackslash}p{2.5cm}>{\raggedleft\arraybackslash}p{2.5cm}}
\toprule
Description & 2005 & 2006 & 2007 & 2008 & 2009\\
\midrule
\textbf{\cellcolor{gray!10}{n}} & \cellcolor{gray!10}{1,243} & \cellcolor{gray!10}{1,072} & \cellcolor{gray!10}{915} & \cellcolor{gray!10}{793} & \cellcolor{gray!10}{681}\\
\textbf{PctNoExpense} & 7.48 & 7.74 & 8.09 & 10.47 & 11.01\\
\textbf{\cellcolor{gray!10}{p25}} & \cellcolor{gray!10}{179} & \cellcolor{gray!10}{160} & \cellcolor{gray!10}{180} & \cellcolor{gray!10}{195} & \cellcolor{gray!10}{177}\\
\textbf{p50} & 406 & 379 & 433 & 450 & 416\\
\textbf{\cellcolor{gray!10}{p75}} & \cellcolor{gray!10}{990} & \cellcolor{gray!10}{858} & \cellcolor{gray!10}{906} & \cellcolor{gray!10}{1,079} & \cellcolor{gray!10}{1,014}\\
\addlinespace
\textbf{p90} & 2,192 & 1,927 & 1,802 & 2,349 & 2,500\\
\textbf{\cellcolor{gray!10}{p95}} & \cellcolor{gray!10}{3,635} & \cellcolor{gray!10}{4,106} & \cellcolor{gray!10}{3,435} & \cellcolor{gray!10}{4,680} & \cellcolor{gray!10}{4,462}\\
\textbf{p96} & 4,197 & 5,306 & 3,999 & 6,085 & 5,243\\
\textbf{\cellcolor{gray!10}{p97}} & \cellcolor{gray!10}{6,112} & \cellcolor{gray!10}{7,026} & \cellcolor{gray!10}{5,596} & \cellcolor{gray!10}{6,553} & \cellcolor{gray!10}{6,861}\\
\textbf{p98} & 9,181 & 9,000 & 7,460 & 11,110 & 10,335\\
\addlinespace
\textbf{\cellcolor{gray!10}{p99}} & \cellcolor{gray!10}{16,453} & \cellcolor{gray!10}{18,774} & \cellcolor{gray!10}{13,060} & \cellcolor{gray!10}{20,153} & \cellcolor{gray!10}{19,082}\\
\textbf{p995} & 26,250 & 21,917 & 19,561 & 35,258 & 54,969\\
\textbf{\cellcolor{gray!10}{p999}} & \cellcolor{gray!10}{63,068} & \cellcolor{gray!10}{67,481} & \cellcolor{gray!10}{66,575} & \cellcolor{gray!10}{95,351} & \cellcolor{gray!10}{201,729}\\
\textbf{max} & 229,290 & 128,291 & 199,011 & 147,129 & 323,604\\
\textbf{\cellcolor{gray!10}{mean}} & \cellcolor{gray!10}{1,388} & \cellcolor{gray!10}{1,290} & \cellcolor{gray!10}{1,255} & \cellcolor{gray!10}{1,578} & \cellcolor{gray!10}{2,038}\\
\addlinespace
\textbf{sd} & 7,579 & 5,432 & 7,288 & 7,035 & 14,706\\
\bottomrule
\end{tabular}
\label{tab:table3.4}
\end{table}
\endgroup
\textbf{n}: total of individuals in portfolio;\textbf{p(k)}: k-th sample percentile;\textbf{SD}: standard deviation

\medskip

% latex table generated in R 4 2.3 by append_kable_to_file function
% Current Date/Time: 2024-02-18 01:18:01
\begingroup
\renewcommand{\thetable}{3.5}
\begin{table}[H]
\centering
\caption{Descriptive Statistics Annual Expenses (Female) 31a35}
\centering
\begin{tabular}[t]{>{}l>{\raggedleft\arraybackslash}p{2.5cm}>{\raggedleft\arraybackslash}p{2.5cm}>{\raggedleft\arraybackslash}p{2.5cm}>{\raggedleft\arraybackslash}p{2.5cm}>{\raggedleft\arraybackslash}p{2.5cm}}
\toprule
Description & 2005 & 2006 & 2007 & 2008 & 2009\\
\midrule
\textbf{\cellcolor{gray!10}{n}} & \cellcolor{gray!10}{1,872} & \cellcolor{gray!10}{1,806} & \cellcolor{gray!10}{1,707} & \cellcolor{gray!10}{1,566} & \cellcolor{gray!10}{1,454}\\
\textbf{PctNoExpense} & 5.02 & 4.32 & 4.98 & 4.92 & 5.85\\
\textbf{\cellcolor{gray!10}{p25}} & \cellcolor{gray!10}{368} & \cellcolor{gray!10}{335} & \cellcolor{gray!10}{393} & \cellcolor{gray!10}{373} & \cellcolor{gray!10}{341}\\
\textbf{p50} & 845 & 817 & 863 & 942 & 804\\
\textbf{\cellcolor{gray!10}{p75}} & \cellcolor{gray!10}{1,948} & \cellcolor{gray!10}{1,884} & \cellcolor{gray!10}{1,909} & \cellcolor{gray!10}{2,245} & \cellcolor{gray!10}{1,877}\\
\addlinespace
\textbf{p90} & 4,659 & 4,735 & 4,757 & 5,259 & 4,887\\
\textbf{\cellcolor{gray!10}{p95}} & \cellcolor{gray!10}{8,907} & \cellcolor{gray!10}{7,946} & \cellcolor{gray!10}{8,319} & \cellcolor{gray!10}{9,141} & \cellcolor{gray!10}{8,771}\\
\textbf{p96} & 10,201 & 9,012 & 9,227 & 10,004 & 9,550\\
\textbf{\cellcolor{gray!10}{p97}} & \cellcolor{gray!10}{11,406} & \cellcolor{gray!10}{9,878} & \cellcolor{gray!10}{10,565} & \cellcolor{gray!10}{10,924} & \cellcolor{gray!10}{10,444}\\
\textbf{p98} & 13,272 & 10,757 & 13,401 & 13,968 & 14,317\\
\addlinespace
\textbf{\cellcolor{gray!10}{p99}} & \cellcolor{gray!10}{17,819} & \cellcolor{gray!10}{15,736} & \cellcolor{gray!10}{20,232} & \cellcolor{gray!10}{18,703} & \cellcolor{gray!10}{19,230}\\
\textbf{p995} & 28,856 & 24,549 & 31,337 & 27,130 & 33,526\\
\textbf{\cellcolor{gray!10}{p999}} & \cellcolor{gray!10}{59,213} & \cellcolor{gray!10}{64,758} & \cellcolor{gray!10}{70,975} & \cellcolor{gray!10}{71,472} & \cellcolor{gray!10}{52,075}\\
\textbf{max} & 87,084 & 101,389 & 98,995 & 141,717 & 69,533\\
\textbf{\cellcolor{gray!10}{mean}} & \cellcolor{gray!10}{2,146} & \cellcolor{gray!10}{1,964} & \cellcolor{gray!10}{2,174} & \cellcolor{gray!10}{2,342} & \cellcolor{gray!10}{2,101}\\
\addlinespace
\textbf{sd} & 4,815 & 4,583 & 5,234 & 5,756 & 4,532\\
\bottomrule
\end{tabular}
\label{tab:table3.5}
\end{table}
\endgroup
\textbf{n}: total of individuals in portfolio;\textbf{p(k)}: k-th sample percentile;\textbf{SD}: standard deviation

\medskip

% latex table generated in R 4 2.3 by append_kable_to_file function
% Current Date/Time: 2024-02-18 01:18:02
\begingroup
\renewcommand{\thetable}{3.6}
\begin{table}[H]
\centering
\caption{Descriptive Statistics Annual Expenses (Male) 31a35}
\centering
\begin{tabular}[t]{>{}l>{\raggedleft\arraybackslash}p{2.5cm}>{\raggedleft\arraybackslash}p{2.5cm}>{\raggedleft\arraybackslash}p{2.5cm}>{\raggedleft\arraybackslash}p{2.5cm}>{\raggedleft\arraybackslash}p{2.5cm}}
\toprule
Description & 2005 & 2006 & 2007 & 2008 & 2009\\
\midrule
\textbf{\cellcolor{gray!10}{n}} & \cellcolor{gray!10}{1,908} & \cellcolor{gray!10}{1,801} & \cellcolor{gray!10}{1,664} & \cellcolor{gray!10}{1,462} & \cellcolor{gray!10}{1,248}\\
\textbf{PctNoExpense} & 7.39 & 6.89 & 8.17 & 7.25 & 7.05\\
\textbf{\cellcolor{gray!10}{p25}} & \cellcolor{gray!10}{195} & \cellcolor{gray!10}{181} & \cellcolor{gray!10}{188} & \cellcolor{gray!10}{206} & \cellcolor{gray!10}{176}\\
\textbf{p50} & 460 & 436 & 475 & 527 & 483\\
\textbf{\cellcolor{gray!10}{p75}} & \cellcolor{gray!10}{1,026} & \cellcolor{gray!10}{976} & \cellcolor{gray!10}{1,106} & \cellcolor{gray!10}{1,205} & \cellcolor{gray!10}{1,123}\\
\addlinespace
\textbf{p90} & 2,010 & 2,044 & 2,342 & 2,862 & 2,561\\
\textbf{\cellcolor{gray!10}{p95}} & \cellcolor{gray!10}{3,546} & \cellcolor{gray!10}{3,299} & \cellcolor{gray!10}{4,449} & \cellcolor{gray!10}{5,792} & \cellcolor{gray!10}{4,816}\\
\textbf{p96} & 4,413 & 3,778 & 5,336 & 8,042 & 6,489\\
\textbf{\cellcolor{gray!10}{p97}} & \cellcolor{gray!10}{5,803} & \cellcolor{gray!10}{4,659} & \cellcolor{gray!10}{6,675} & \cellcolor{gray!10}{10,330} & \cellcolor{gray!10}{8,563}\\
\textbf{p98} & 8,308 & 7,118 & 9,835 & 12,979 & 12,292\\
\addlinespace
\textbf{\cellcolor{gray!10}{p99}} & \cellcolor{gray!10}{13,512} & \cellcolor{gray!10}{15,364} & \cellcolor{gray!10}{19,401} & \cellcolor{gray!10}{22,030} & \cellcolor{gray!10}{18,377}\\
\textbf{p995} & 22,473 & 21,891 & 25,576 & 42,934 & 21,802\\
\textbf{\cellcolor{gray!10}{p999}} & \cellcolor{gray!10}{67,187} & \cellcolor{gray!10}{107,083} & \cellcolor{gray!10}{63,819} & \cellcolor{gray!10}{92,329} & \cellcolor{gray!10}{42,142}\\
\textbf{max} & 130,614 & 172,207 & 90,771 & 113,230 & 61,267\\
\textbf{\cellcolor{gray!10}{mean}} & \cellcolor{gray!10}{1,243} & \cellcolor{gray!10}{1,295} & \cellcolor{gray!10}{1,357} & \cellcolor{gray!10}{1,792} & \cellcolor{gray!10}{1,396}\\
\addlinespace
\textbf{sd} & 4,771 & 6,417 & 4,446 & 6,534 & 3,639\\
\bottomrule
\end{tabular}
\label{tab:table3.6}
\end{table}
\endgroup
\textbf{n}: total of individuals in portfolio;\textbf{p(k)}: k-th sample percentile;\textbf{SD}: standard deviation

\medskip

% latex table generated in R 4 2.3 by append_kable_to_file function
% Current Date/Time: 2024-02-18 01:18:02
\begingroup
\renewcommand{\thetable}{3.7}
\begin{table}[H]
\centering
\caption{Descriptive Statistics Annual Expenses (Female) 36a40}
\centering
\begin{tabular}[t]{>{}l>{\raggedleft\arraybackslash}p{2.5cm}>{\raggedleft\arraybackslash}p{2.5cm}>{\raggedleft\arraybackslash}p{2.5cm}>{\raggedleft\arraybackslash}p{2.5cm}>{\raggedleft\arraybackslash}p{2.5cm}}
\toprule
Description & 2005 & 2006 & 2007 & 2008 & 2009\\
\midrule
\textbf{\cellcolor{gray!10}{n}} & \cellcolor{gray!10}{2,242} & \cellcolor{gray!10}{2,204} & \cellcolor{gray!10}{2,126} & \cellcolor{gray!10}{2,056} & \cellcolor{gray!10}{1,959}\\
\textbf{PctNoExpense} & 5.62 & 6.72 & 6.40 & 5.54 & 5.77\\
\textbf{\cellcolor{gray!10}{p25}} & \cellcolor{gray!10}{363} & \cellcolor{gray!10}{344} & \cellcolor{gray!10}{346} & \cellcolor{gray!10}{379} & \cellcolor{gray!10}{390}\\
\textbf{p50} & 793 & 762 & 791 & 916 & 937\\
\textbf{\cellcolor{gray!10}{p75}} & \cellcolor{gray!10}{1,871} & \cellcolor{gray!10}{1,628} & \cellcolor{gray!10}{1,802} & \cellcolor{gray!10}{2,141} & \cellcolor{gray!10}{2,053}\\
\addlinespace
\textbf{p90} & 4,520 & 3,644 & 4,375 & 5,012 & 4,378\\
\textbf{\cellcolor{gray!10}{p95}} & \cellcolor{gray!10}{8,043} & \cellcolor{gray!10}{6,276} & \cellcolor{gray!10}{7,735} & \cellcolor{gray!10}{8,502} & \cellcolor{gray!10}{8,314}\\
\textbf{p96} & 9,237 & 7,244 & 8,981 & 9,400 & 9,402\\
\textbf{\cellcolor{gray!10}{p97}} & \cellcolor{gray!10}{10,519} & \cellcolor{gray!10}{8,840} & \cellcolor{gray!10}{10,607} & \cellcolor{gray!10}{10,728} & \cellcolor{gray!10}{10,757}\\
\textbf{p98} & 12,118 & 10,912 & 13,138 & 13,291 & 13,554\\
\addlinespace
\textbf{\cellcolor{gray!10}{p99}} & \cellcolor{gray!10}{18,267} & \cellcolor{gray!10}{15,634} & \cellcolor{gray!10}{20,821} & \cellcolor{gray!10}{21,374} & \cellcolor{gray!10}{19,857}\\
\textbf{p995} & 25,417 & 31,130 & 56,363 & 40,889 & 27,142\\
\textbf{\cellcolor{gray!10}{p999}} & \cellcolor{gray!10}{75,004} & \cellcolor{gray!10}{102,143} & \cellcolor{gray!10}{85,783} & \cellcolor{gray!10}{176,175} & \cellcolor{gray!10}{113,373}\\
\textbf{max} & 157,006 & 528,293 & 92,919 & 318,932 & 221,151\\
\textbf{\cellcolor{gray!10}{mean}} & \cellcolor{gray!10}{2,058} & \cellcolor{gray!10}{2,154} & \cellcolor{gray!10}{2,153} & \cellcolor{gray!10}{2,745} & \cellcolor{gray!10}{2,336}\\
\addlinespace
\textbf{sd} & 5,657 & 13,605 & 6,006 & 12,851 & 8,471\\
\bottomrule
\end{tabular}
\label{tab:table3.7}
\end{table}
\endgroup
\textbf{n}: total of individuals in portfolio;\textbf{p(k)}: k-th sample percentile;\textbf{SD}: standard deviation

\medskip

% latex table generated in R 4 2.3 by append_kable_to_file function
% Current Date/Time: 2024-02-18 01:18:02
\begingroup
\renewcommand{\thetable}{3.8}
\begin{table}[H]
\centering
\caption{Descriptive Statistics Annual Expenses (Male) 36a40}
\centering
\begin{tabular}[t]{>{}l>{\raggedleft\arraybackslash}p{2.5cm}>{\raggedleft\arraybackslash}p{2.5cm}>{\raggedleft\arraybackslash}p{2.5cm}>{\raggedleft\arraybackslash}p{2.5cm}>{\raggedleft\arraybackslash}p{2.5cm}}
\toprule
Description & 2005 & 2006 & 2007 & 2008 & 2009\\
\midrule
\textbf{\cellcolor{gray!10}{n}} & \cellcolor{gray!10}{2,286} & \cellcolor{gray!10}{2,195} & \cellcolor{gray!10}{2,139} & \cellcolor{gray!10}{2,088} & \cellcolor{gray!10}{2,052}\\
\textbf{PctNoExpense} & 7.09 & 6.33 & 6.69 & 6.90 & 7.26\\
\textbf{\cellcolor{gray!10}{p25}} & \cellcolor{gray!10}{209} & \cellcolor{gray!10}{216} & \cellcolor{gray!10}{214} & \cellcolor{gray!10}{229} & \cellcolor{gray!10}{204}\\
\textbf{p50} & 513 & 496 & 500 & 553 & 512\\
\textbf{\cellcolor{gray!10}{p75}} & \cellcolor{gray!10}{1,211} & \cellcolor{gray!10}{1,111} & \cellcolor{gray!10}{1,098} & \cellcolor{gray!10}{1,299} & \cellcolor{gray!10}{1,170}\\
\addlinespace
\textbf{p90} & 2,682 & 2,424 & 2,386 & 2,679 & 2,744\\
\textbf{\cellcolor{gray!10}{p95}} & \cellcolor{gray!10}{5,035} & \cellcolor{gray!10}{4,506} & \cellcolor{gray!10}{4,694} & \cellcolor{gray!10}{4,616} & \cellcolor{gray!10}{5,537}\\
\textbf{p96} & 6,375 & 5,409 & 5,287 & 5,164 & 6,540\\
\textbf{\cellcolor{gray!10}{p97}} & \cellcolor{gray!10}{7,534} & \cellcolor{gray!10}{6,514} & \cellcolor{gray!10}{6,862} & \cellcolor{gray!10}{6,745} & \cellcolor{gray!10}{8,344}\\
\textbf{p98} & 10,579 & 9,115 & 9,601 & 9,955 & 11,419\\
\addlinespace
\textbf{\cellcolor{gray!10}{p99}} & \cellcolor{gray!10}{19,902} & \cellcolor{gray!10}{14,195} & \cellcolor{gray!10}{17,052} & \cellcolor{gray!10}{16,145} & \cellcolor{gray!10}{20,056}\\
\textbf{p995} & 24,752 & 21,963 & 34,069 & 22,558 & 31,768\\
\textbf{\cellcolor{gray!10}{p999}} & \cellcolor{gray!10}{99,053} & \cellcolor{gray!10}{42,568} & \cellcolor{gray!10}{83,016} & \cellcolor{gray!10}{71,215} & \cellcolor{gray!10}{115,938}\\
\textbf{max} & 230,437 & 162,392 & 211,526 & 286,716 & 142,242\\
\textbf{\cellcolor{gray!10}{mean}} & \cellcolor{gray!10}{1,564} & \cellcolor{gray!10}{1,327} & \cellcolor{gray!10}{1,535} & \cellcolor{gray!10}{1,558} & \cellcolor{gray!10}{1,635}\\
\addlinespace
\textbf{sd} & 6,815 & 4,665 & 6,983 & 7,595 & 6,326\\
\bottomrule
\end{tabular}
\label{tab:table3.8}
\end{table}
\endgroup
\textbf{n}: total of individuals in portfolio;\textbf{p(k)}: k-th sample percentile;\textbf{SD}: standard deviation

\medskip

% latex table generated in R 4 2.3 by append_kable_to_file function
% Current Date/Time: 2024-02-18 01:18:02
\begingroup
\renewcommand{\thetable}{3.9}
\begin{table}[H]
\centering
\caption{Descriptive Statistics Annual Expenses (Female) 41a45}
\centering
\begin{tabular}[t]{>{}l>{\raggedleft\arraybackslash}p{2.5cm}>{\raggedleft\arraybackslash}p{2.5cm}>{\raggedleft\arraybackslash}p{2.5cm}>{\raggedleft\arraybackslash}p{2.5cm}>{\raggedleft\arraybackslash}p{2.5cm}}
\toprule
Description & 2005 & 2006 & 2007 & 2008 & 2009\\
\midrule
\textbf{\cellcolor{gray!10}{n}} & \cellcolor{gray!10}{2,557} & \cellcolor{gray!10}{2,494} & \cellcolor{gray!10}{2,454} & \cellcolor{gray!10}{2,382} & \cellcolor{gray!10}{2,339}\\
\textbf{PctNoExpense} & 6.14 & 6.01 & 5.30 & 5.58 & 6.67\\
\textbf{\cellcolor{gray!10}{p25}} & \cellcolor{gray!10}{455} & \cellcolor{gray!10}{413} & \cellcolor{gray!10}{434} & \cellcolor{gray!10}{443} & \cellcolor{gray!10}{447}\\
\textbf{p50} & 949 & 859 & 912 & 985 & 977\\
\textbf{\cellcolor{gray!10}{p75}} & \cellcolor{gray!10}{2,031} & \cellcolor{gray!10}{1,804} & \cellcolor{gray!10}{1,887} & \cellcolor{gray!10}{2,125} & \cellcolor{gray!10}{2,021}\\
\addlinespace
\textbf{p90} & 4,369 & 3,735 & 4,018 & 4,842 & 4,648\\
\textbf{\cellcolor{gray!10}{p95}} & \cellcolor{gray!10}{7,502} & \cellcolor{gray!10}{6,142} & \cellcolor{gray!10}{6,514} & \cellcolor{gray!10}{8,198} & \cellcolor{gray!10}{8,096}\\
\textbf{p96} & 8,718 & 7,183 & 7,572 & 9,834 & 9,821\\
\textbf{\cellcolor{gray!10}{p97}} & \cellcolor{gray!10}{10,362} & \cellcolor{gray!10}{9,017} & \cellcolor{gray!10}{9,128} & \cellcolor{gray!10}{11,537} & \cellcolor{gray!10}{10,956}\\
\textbf{p98} & 13,856 & 11,100 & 12,013 & 14,644 & 13,852\\
\addlinespace
\textbf{\cellcolor{gray!10}{p99}} & \cellcolor{gray!10}{25,158} & \cellcolor{gray!10}{18,653} & \cellcolor{gray!10}{19,821} & \cellcolor{gray!10}{23,148} & \cellcolor{gray!10}{21,052}\\
\textbf{p995} & 34,825 & 30,076 & 28,599 & 35,298 & 50,353\\
\textbf{\cellcolor{gray!10}{p999}} & \cellcolor{gray!10}{48,097} & \cellcolor{gray!10}{81,079} & \cellcolor{gray!10}{115,517} & \cellcolor{gray!10}{118,538} & \cellcolor{gray!10}{177,314}\\
\textbf{max} & 66,973 & 162,119 & 196,400 & 201,497 & 254,654\\
\textbf{\cellcolor{gray!10}{mean}} & \cellcolor{gray!10}{2,140} & \cellcolor{gray!10}{1,993} & \cellcolor{gray!10}{2,188} & \cellcolor{gray!10}{2,503} & \cellcolor{gray!10}{2,563}\\
\addlinespace
\textbf{sd} & 4,509 & 6,006 & 7,444 & 8,070 & 9,987\\
\bottomrule
\end{tabular}
\label{tab:table3.9}
\end{table}
\endgroup
\textbf{n}: total of individuals in portfolio;\textbf{p(k)}: k-th sample percentile;\textbf{SD}: standard deviation

\medskip

% latex table generated in R 4 2.3 by append_kable_to_file function
% Current Date/Time: 2024-02-18 01:18:03
\begingroup
\renewcommand{\thetable}{3.10}
\begin{table}[H]
\centering
\caption{Descriptive Statistics Annual Expenses (Male) 41a45}
\centering
\begin{tabular}[t]{>{}l>{\raggedleft\arraybackslash}p{2.5cm}>{\raggedleft\arraybackslash}p{2.5cm}>{\raggedleft\arraybackslash}p{2.5cm}>{\raggedleft\arraybackslash}p{2.5cm}>{\raggedleft\arraybackslash}p{2.5cm}}
\toprule
Description & 2005 & 2006 & 2007 & 2008 & 2009\\
\midrule
\textbf{\cellcolor{gray!10}{n}} & \cellcolor{gray!10}{2,489} & \cellcolor{gray!10}{2,446} & \cellcolor{gray!10}{2,376} & \cellcolor{gray!10}{2,381} & \cellcolor{gray!10}{2,317}\\
\textbf{PctNoExpense} & 7.35 & 5.27 & 6.02 & 5.50 & 4.92\\
\textbf{\cellcolor{gray!10}{p25}} & \cellcolor{gray!10}{233} & \cellcolor{gray!10}{225} & \cellcolor{gray!10}{230} & \cellcolor{gray!10}{270} & \cellcolor{gray!10}{246}\\
\textbf{p50} & 559 & 534 & 566 & 673 & 573\\
\textbf{\cellcolor{gray!10}{p75}} & \cellcolor{gray!10}{1,312} & \cellcolor{gray!10}{1,272} & \cellcolor{gray!10}{1,280} & \cellcolor{gray!10}{1,498} & \cellcolor{gray!10}{1,306}\\
\addlinespace
\textbf{p90} & 2,765 & 2,692 & 2,832 & 3,318 & 2,987\\
\textbf{\cellcolor{gray!10}{p95}} & \cellcolor{gray!10}{4,860} & \cellcolor{gray!10}{5,530} & \cellcolor{gray!10}{5,428} & \cellcolor{gray!10}{6,306} & \cellcolor{gray!10}{6,220}\\
\textbf{p96} & 5,590 & 6,698 & 6,697 & 7,610 & 7,884\\
\textbf{\cellcolor{gray!10}{p97}} & \cellcolor{gray!10}{7,162} & \cellcolor{gray!10}{8,705} & \cellcolor{gray!10}{8,475} & \cellcolor{gray!10}{9,842} & \cellcolor{gray!10}{9,736}\\
\textbf{p98} & 10,509 & 11,371 & 12,263 & 14,657 & 13,208\\
\addlinespace
\textbf{\cellcolor{gray!10}{p99}} & \cellcolor{gray!10}{17,289} & \cellcolor{gray!10}{16,961} & \cellcolor{gray!10}{19,340} & \cellcolor{gray!10}{25,622} & \cellcolor{gray!10}{24,125}\\
\textbf{p995} & 32,462 & 25,310 & 39,384 & 40,011 & 32,908\\
\textbf{\cellcolor{gray!10}{p999}} & \cellcolor{gray!10}{91,013} & \cellcolor{gray!10}{54,602} & \cellcolor{gray!10}{71,849} & \cellcolor{gray!10}{94,264} & \cellcolor{gray!10}{75,851}\\
\textbf{max} & 117,357 & 267,607 & 244,532 & 140,615 & 114,218\\
\textbf{\cellcolor{gray!10}{mean}} & \cellcolor{gray!10}{1,584} & \cellcolor{gray!10}{1,600} & \cellcolor{gray!10}{1,718} & \cellcolor{gray!10}{2,005} & \cellcolor{gray!10}{1,735}\\
\addlinespace
\textbf{sd} & 5,447 & 6,817 & 7,229 & 7,051 & 5,285\\
\bottomrule
\end{tabular}
\label{tab:table3.10}
\end{table}
\endgroup
\textbf{n}: total of individuals in portfolio;\textbf{p(k)}: k-th sample percentile;\textbf{SD}: standard deviation

\medskip

% latex table generated in R 4 2.3 by append_kable_to_file function
% Current Date/Time: 2024-02-18 01:18:03
\begingroup
\renewcommand{\thetable}{3.11}
\begin{table}[H]
\centering
\caption{Descriptive Statistics Annual Expenses (Female) 46a50}
\centering
\begin{tabular}[t]{>{}l>{\raggedleft\arraybackslash}p{2.5cm}>{\raggedleft\arraybackslash}p{2.5cm}>{\raggedleft\arraybackslash}p{2.5cm}>{\raggedleft\arraybackslash}p{2.5cm}>{\raggedleft\arraybackslash}p{2.5cm}}
\toprule
Description & 2005 & 2006 & 2007 & 2008 & 2009\\
\midrule
\textbf{\cellcolor{gray!10}{n}} & \cellcolor{gray!10}{2,337} & \cellcolor{gray!10}{2,440} & \cellcolor{gray!10}{2,464} & \cellcolor{gray!10}{2,520} & \cellcolor{gray!10}{2,531}\\
\textbf{PctNoExpense} & 4.45 & 5.00 & 5.24 & 5.40 & 5.33\\
\textbf{\cellcolor{gray!10}{p25}} & \cellcolor{gray!10}{561} & \cellcolor{gray!10}{507} & \cellcolor{gray!10}{545} & \cellcolor{gray!10}{543} & \cellcolor{gray!10}{532}\\
\textbf{p50} & 1,131 & 1,058 & 1,121 & 1,234 & 1,133\\
\textbf{\cellcolor{gray!10}{p75}} & \cellcolor{gray!10}{2,225} & \cellcolor{gray!10}{2,186} & \cellcolor{gray!10}{2,259} & \cellcolor{gray!10}{2,588} & \cellcolor{gray!10}{2,473}\\
\addlinespace
\textbf{p90} & 4,625 & 4,572 & 4,471 & 5,128 & 5,336\\
\textbf{\cellcolor{gray!10}{p95}} & \cellcolor{gray!10}{7,772} & \cellcolor{gray!10}{7,324} & \cellcolor{gray!10}{8,005} & \cellcolor{gray!10}{8,436} & \cellcolor{gray!10}{9,916}\\
\textbf{p96} & 9,564 & 8,366 & 9,267 & 9,812 & 12,467\\
\textbf{\cellcolor{gray!10}{p97}} & \cellcolor{gray!10}{11,216} & \cellcolor{gray!10}{10,515} & \cellcolor{gray!10}{11,043} & \cellcolor{gray!10}{11,603} & \cellcolor{gray!10}{14,748}\\
\textbf{p98} & 14,077 & 13,077 & 14,062 & 15,898 & 17,793\\
\addlinespace
\textbf{\cellcolor{gray!10}{p99}} & \cellcolor{gray!10}{24,609} & \cellcolor{gray!10}{24,373} & \cellcolor{gray!10}{22,171} & \cellcolor{gray!10}{25,993} & \cellcolor{gray!10}{25,841}\\
\textbf{p995} & 27,863 & 37,853 & 35,379 & 55,079 & 45,159\\
\textbf{\cellcolor{gray!10}{p999}} & \cellcolor{gray!10}{50,564} & \cellcolor{gray!10}{70,935} & \cellcolor{gray!10}{70,728} & \cellcolor{gray!10}{119,113} & \cellcolor{gray!10}{129,367}\\
\textbf{max} & 89,188 & 177,710 & 207,664 & 203,918 & 348,062\\
\textbf{\cellcolor{gray!10}{mean}} & \cellcolor{gray!10}{2,296} & \cellcolor{gray!10}{2,322} & \cellcolor{gray!10}{2,434} & \cellcolor{gray!10}{2,795} & \cellcolor{gray!10}{2,946}\\
\addlinespace
\textbf{sd} & 4,482 & 6,132 & 6,664 & 8,123 & 10,707\\
\bottomrule
\end{tabular}
\label{tab:table3.11}
\end{table}
\endgroup
\textbf{n}: total of individuals in portfolio;\textbf{p(k)}: k-th sample percentile;\textbf{SD}: standard deviation

\medskip

% latex table generated in R 4 2.3 by append_kable_to_file function
% Current Date/Time: 2024-02-18 01:18:03
\begingroup
\renewcommand{\thetable}{3.12}
\begin{table}[H]
\centering
\caption{Descriptive Statistics Annual Expenses (Male) 46a50}
\centering
\begin{tabular}[t]{>{}l>{\raggedleft\arraybackslash}p{2.5cm}>{\raggedleft\arraybackslash}p{2.5cm}>{\raggedleft\arraybackslash}p{2.5cm}>{\raggedleft\arraybackslash}p{2.5cm}>{\raggedleft\arraybackslash}p{2.5cm}}
\toprule
Description & 2005 & 2006 & 2007 & 2008 & 2009\\
\midrule
\textbf{\cellcolor{gray!10}{n}} & \cellcolor{gray!10}{2,905} & \cellcolor{gray!10}{2,855} & \cellcolor{gray!10}{2,767} & \cellcolor{gray!10}{2,625} & \cellcolor{gray!10}{2,515}\\
\textbf{PctNoExpense} & 5.61 & 4.59 & 5.35 & 5.22 & 5.61\\
\textbf{\cellcolor{gray!10}{p25}} & \cellcolor{gray!10}{263} & \cellcolor{gray!10}{257} & \cellcolor{gray!10}{281} & \cellcolor{gray!10}{299} & \cellcolor{gray!10}{281}\\
\textbf{p50} & 646 & 617 & 674 & 740 & 680\\
\textbf{\cellcolor{gray!10}{p75}} & \cellcolor{gray!10}{1,439} & \cellcolor{gray!10}{1,440} & \cellcolor{gray!10}{1,477} & \cellcolor{gray!10}{1,661} & \cellcolor{gray!10}{1,525}\\
\addlinespace
\textbf{p90} & 3,139 & 3,108 & 3,250 & 3,873 & 3,545\\
\textbf{\cellcolor{gray!10}{p95}} & \cellcolor{gray!10}{5,968} & \cellcolor{gray!10}{5,723} & \cellcolor{gray!10}{6,672} & \cellcolor{gray!10}{7,680} & \cellcolor{gray!10}{8,102}\\
\textbf{p96} & 7,164 & 7,470 & 8,149 & 9,585 & 10,445\\
\textbf{\cellcolor{gray!10}{p97}} & \cellcolor{gray!10}{9,182} & \cellcolor{gray!10}{9,022} & \cellcolor{gray!10}{11,134} & \cellcolor{gray!10}{15,316} & \cellcolor{gray!10}{15,606}\\
\textbf{p98} & 13,455 & 12,634 & 15,129 & 23,472 & 22,146\\
\addlinespace
\textbf{\cellcolor{gray!10}{p99}} & \cellcolor{gray!10}{20,150} & \cellcolor{gray!10}{23,003} & \cellcolor{gray!10}{27,865} & \cellcolor{gray!10}{37,801} & \cellcolor{gray!10}{45,070}\\
\textbf{p995} & 30,414 & 35,122 & 48,033 & 62,346 & 70,808\\
\textbf{\cellcolor{gray!10}{p999}} & \cellcolor{gray!10}{98,447} & \cellcolor{gray!10}{86,831} & \cellcolor{gray!10}{89,023} & \cellcolor{gray!10}{113,012} & \cellcolor{gray!10}{160,275}\\
\textbf{max} & 426,772 & 137,333 & 124,276 & 275,813 & 397,015\\
\textbf{\cellcolor{gray!10}{mean}} & \cellcolor{gray!10}{1,921} & \cellcolor{gray!10}{1,808} & \cellcolor{gray!10}{2,005} & \cellcolor{gray!10}{2,585} & \cellcolor{gray!10}{2,747}\\
\addlinespace
\textbf{sd} & 9,858 & 5,929 & 6,495 & 10,411 & 13,107\\
\bottomrule
\end{tabular}
\label{tab:table3.12}
\end{table}
\endgroup
\textbf{n}: total of individuals in portfolio;\textbf{p(k)}: k-th sample percentile;\textbf{SD}: standard deviation

\medskip

% latex table generated in R 4 2.3 by append_kable_to_file function
% Current Date/Time: 2024-02-18 01:18:04
\begingroup
\renewcommand{\thetable}{3.13}
\begin{table}[H]
\centering
\caption{Descriptive Statistics Annual Expenses (Female) 51a55}
\centering
\begin{tabular}[t]{>{}l>{\raggedleft\arraybackslash}p{2.5cm}>{\raggedleft\arraybackslash}p{2.5cm}>{\raggedleft\arraybackslash}p{2.5cm}>{\raggedleft\arraybackslash}p{2.5cm}>{\raggedleft\arraybackslash}p{2.5cm}}
\toprule
Description & 2005 & 2006 & 2007 & 2008 & 2009\\
\midrule
\textbf{\cellcolor{gray!10}{n}} & \cellcolor{gray!10}{1,491} & \cellcolor{gray!10}{1,646} & \cellcolor{gray!10}{1,847} & \cellcolor{gray!10}{2,031} & \cellcolor{gray!10}{2,204}\\
\textbf{PctNoExpense} & 4.63 & 3.65 & 4.44 & 4.38 & 4.95\\
\textbf{\cellcolor{gray!10}{p25}} & \cellcolor{gray!10}{703} & \cellcolor{gray!10}{656} & \cellcolor{gray!10}{584} & \cellcolor{gray!10}{656} & \cellcolor{gray!10}{620}\\
\textbf{p50} & 1,394 & 1,265 & 1,207 & 1,388 & 1,337\\
\textbf{\cellcolor{gray!10}{p75}} & \cellcolor{gray!10}{2,625} & \cellcolor{gray!10}{2,517} & \cellcolor{gray!10}{2,547} & \cellcolor{gray!10}{2,880} & \cellcolor{gray!10}{2,900}\\
\addlinespace
\textbf{p90} & 5,317 & 5,024 & 5,686 & 6,299 & 6,104\\
\textbf{\cellcolor{gray!10}{p95}} & \cellcolor{gray!10}{9,427} & \cellcolor{gray!10}{8,570} & \cellcolor{gray!10}{9,812} & \cellcolor{gray!10}{10,487} & \cellcolor{gray!10}{10,003}\\
\textbf{p96} & 11,851 & 10,229 & 11,274 & 12,000 & 12,111\\
\textbf{\cellcolor{gray!10}{p97}} & \cellcolor{gray!10}{15,409} & \cellcolor{gray!10}{12,954} & \cellcolor{gray!10}{14,602} & \cellcolor{gray!10}{13,995} & \cellcolor{gray!10}{17,558}\\
\textbf{p98} & 20,293 & 19,269 & 19,709 & 21,897 & 27,194\\
\addlinespace
\textbf{\cellcolor{gray!10}{p99}} & \cellcolor{gray!10}{34,516} & \cellcolor{gray!10}{34,645} & \cellcolor{gray!10}{35,600} & \cellcolor{gray!10}{43,418} & \cellcolor{gray!10}{43,935}\\
\textbf{p995} & 56,503 & 78,735 & 55,339 & 53,830 & 75,109\\
\textbf{\cellcolor{gray!10}{p999}} & \cellcolor{gray!10}{73,357} & \cellcolor{gray!10}{197,347} & \cellcolor{gray!10}{137,637} & \cellcolor{gray!10}{202,005} & \cellcolor{gray!10}{204,799}\\
\textbf{max} & 282,993 & 297,188 & 355,220 & 1,044,525 & 279,515\\
\textbf{\cellcolor{gray!10}{mean}} & \cellcolor{gray!10}{3,130} & \cellcolor{gray!10}{3,355} & \cellcolor{gray!10}{3,188} & \cellcolor{gray!10}{3,859} & \cellcolor{gray!10}{3,680}\\
\addlinespace
\textbf{sd} & 9,842 & 13,890 & 12,160 & 25,844 & 12,885\\
\bottomrule
\end{tabular}
\label{tab:table3.13}
\end{table}
\endgroup
\textbf{n}: total of individuals in portfolio;\textbf{p(k)}: k-th sample percentile;\textbf{SD}: standard deviation

\medskip

% latex table generated in R 4 2.3 by append_kable_to_file function
% Current Date/Time: 2024-02-18 01:18:04
\begingroup
\renewcommand{\thetable}{3.14}
\begin{table}[H]
\centering
\caption{Descriptive Statistics Annual Expenses (Male) 51a55}
\centering
\begin{tabular}[t]{>{}l>{\raggedleft\arraybackslash}p{2.5cm}>{\raggedleft\arraybackslash}p{2.5cm}>{\raggedleft\arraybackslash}p{2.5cm}>{\raggedleft\arraybackslash}p{2.5cm}>{\raggedleft\arraybackslash}p{2.5cm}}
\toprule
Description & 2005 & 2006 & 2007 & 2008 & 2009\\
\midrule
\textbf{\cellcolor{gray!10}{n}} & \cellcolor{gray!10}{1,972} & \cellcolor{gray!10}{2,263} & \cellcolor{gray!10}{2,552} & \cellcolor{gray!10}{2,755} & \cellcolor{gray!10}{2,928}\\
\textbf{PctNoExpense} & 5.27 & 4.33 & 5.21 & 4.39 & 4.64\\
\textbf{\cellcolor{gray!10}{p25}} & \cellcolor{gray!10}{314} & \cellcolor{gray!10}{312} & \cellcolor{gray!10}{330} & \cellcolor{gray!10}{391} & \cellcolor{gray!10}{345}\\
\textbf{p50} & 750 & 754 & 789 & 884 & 836\\
\textbf{\cellcolor{gray!10}{p75}} & \cellcolor{gray!10}{1,642} & \cellcolor{gray!10}{1,658} & \cellcolor{gray!10}{1,785} & \cellcolor{gray!10}{1,963} & \cellcolor{gray!10}{1,874}\\
\addlinespace
\textbf{p90} & 3,515 & 3,551 & 3,787 & 4,571 & 4,220\\
\textbf{\cellcolor{gray!10}{p95}} & \cellcolor{gray!10}{6,266} & \cellcolor{gray!10}{6,527} & \cellcolor{gray!10}{7,079} & \cellcolor{gray!10}{8,413} & \cellcolor{gray!10}{9,370}\\
\textbf{p96} & 7,223 & 8,288 & 9,252 & 10,396 & 12,245\\
\textbf{\cellcolor{gray!10}{p97}} & \cellcolor{gray!10}{10,201} & \cellcolor{gray!10}{10,694} & \cellcolor{gray!10}{13,255} & \cellcolor{gray!10}{13,862} & \cellcolor{gray!10}{15,997}\\
\textbf{p98} & 13,822 & 15,581 & 20,222 & 23,473 & 24,546\\
\addlinespace
\textbf{\cellcolor{gray!10}{p99}} & \cellcolor{gray!10}{31,181} & \cellcolor{gray!10}{25,131} & \cellcolor{gray!10}{32,155} & \cellcolor{gray!10}{39,463} & \cellcolor{gray!10}{36,614}\\
\textbf{p995} & 58,903 & 40,441 & 44,668 & 67,096 & 51,064\\
\textbf{\cellcolor{gray!10}{p999}} & \cellcolor{gray!10}{109,224} & \cellcolor{gray!10}{100,505} & \cellcolor{gray!10}{90,346} & \cellcolor{gray!10}{157,042} & \cellcolor{gray!10}{171,368}\\
\textbf{max} & 112,453 & 371,231 & 129,112 & 267,595 & 965,287\\
\textbf{\cellcolor{gray!10}{mean}} & \cellcolor{gray!10}{2,177} & \cellcolor{gray!10}{2,221} & \cellcolor{gray!10}{2,273} & \cellcolor{gray!10}{2,893} & \cellcolor{gray!10}{3,027}\\
\addlinespace
\textbf{sd} & 7,391 & 10,533 & 6,786 & 11,132 & 20,648\\
\bottomrule
\end{tabular}
\label{tab:table3.14}
\end{table}
\endgroup
\textbf{n}: total of individuals in portfolio;\textbf{p(k)}: k-th sample percentile;\textbf{SD}: standard deviation

\medskip

% latex table generated in R 4 2.3 by append_kable_to_file function
% Current Date/Time: 2024-02-18 01:18:04
\begingroup
\renewcommand{\thetable}{3.15}
\begin{table}[H]
\centering
\caption{Descriptive Statistics Annual Expenses (Female) 56a60}
\centering
\begin{tabular}[t]{>{}l>{\raggedleft\arraybackslash}p{2.5cm}>{\raggedleft\arraybackslash}p{2.5cm}>{\raggedleft\arraybackslash}p{2.5cm}>{\raggedleft\arraybackslash}p{2.5cm}>{\raggedleft\arraybackslash}p{2.5cm}}
\toprule
Description & 2005 & 2006 & 2007 & 2008 & 2009\\
\midrule
\textbf{\cellcolor{gray!10}{n}} & \cellcolor{gray!10}{831} & \cellcolor{gray!10}{980} & \cellcolor{gray!10}{1,098} & \cellcolor{gray!10}{1,211} & \cellcolor{gray!10}{1,309}\\
\textbf{PctNoExpense} & 6.26 & 6.33 & 5.19 & 4.54 & 3.97\\
\textbf{\cellcolor{gray!10}{p25}} & \cellcolor{gray!10}{627} & \cellcolor{gray!10}{662} & \cellcolor{gray!10}{685} & \cellcolor{gray!10}{780} & \cellcolor{gray!10}{732}\\
\textbf{p50} & 1,366 & 1,346 & 1,432 & 1,615 & 1,648\\
\textbf{\cellcolor{gray!10}{p75}} & \cellcolor{gray!10}{2,563} & \cellcolor{gray!10}{2,705} & \cellcolor{gray!10}{2,863} & \cellcolor{gray!10}{3,477} & \cellcolor{gray!10}{3,520}\\
\addlinespace
\textbf{p90} & 5,487 & 6,278 & 5,818 & 6,573 & 7,741\\
\textbf{\cellcolor{gray!10}{p95}} & \cellcolor{gray!10}{9,145} & \cellcolor{gray!10}{11,310} & \cellcolor{gray!10}{9,959} & \cellcolor{gray!10}{9,957} & \cellcolor{gray!10}{15,414}\\
\textbf{p96} & 11,318 & 13,667 & 11,264 & 11,932 & 19,931\\
\textbf{\cellcolor{gray!10}{p97}} & \cellcolor{gray!10}{14,696} & \cellcolor{gray!10}{17,068} & \cellcolor{gray!10}{17,858} & \cellcolor{gray!10}{16,031} & \cellcolor{gray!10}{24,509}\\
\textbf{p98} & 19,522 & 20,310 & 23,083 & 19,755 & 29,453\\
\addlinespace
\textbf{\cellcolor{gray!10}{p99}} & \cellcolor{gray!10}{28,883} & \cellcolor{gray!10}{27,819} & \cellcolor{gray!10}{29,923} & \cellcolor{gray!10}{37,098} & \cellcolor{gray!10}{51,170}\\
\textbf{p995} & 56,251 & 40,600 & 40,949 & 43,065 & 94,078\\
\textbf{\cellcolor{gray!10}{p999}} & \cellcolor{gray!10}{137,346} & \cellcolor{gray!10}{140,851} & \cellcolor{gray!10}{113,612} & \cellcolor{gray!10}{120,279} & \cellcolor{gray!10}{274,734}\\
\textbf{max} & 209,437 & 222,438 & 148,681 & 340,890 & 550,309\\
\textbf{\cellcolor{gray!10}{mean}} & \cellcolor{gray!10}{3,157} & \cellcolor{gray!10}{3,258} & \cellcolor{gray!10}{3,175} & \cellcolor{gray!10}{3,659} & \cellcolor{gray!10}{4,931}\\
\addlinespace
\textbf{sd} & 10,157 & 9,886 & 8,071 & 12,243 & 21,379\\
\bottomrule
\end{tabular}
\label{tab:table3.15}
\end{table}
\endgroup
\textbf{n}: total of individuals in portfolio;\textbf{p(k)}: k-th sample percentile;\textbf{SD}: standard deviation

\medskip

% latex table generated in R 4 2.3 by append_kable_to_file function
% Current Date/Time: 2024-02-18 01:18:04
\begingroup
\renewcommand{\thetable}{3.16}
\begin{table}[H]
\centering
\caption{Descriptive Statistics Annual Expenses (Male) 56a60}
\centering
\begin{tabular}[t]{>{}l>{\raggedleft\arraybackslash}p{2.5cm}>{\raggedleft\arraybackslash}p{2.5cm}>{\raggedleft\arraybackslash}p{2.5cm}>{\raggedleft\arraybackslash}p{2.5cm}>{\raggedleft\arraybackslash}p{2.5cm}}
\toprule
Description & 2005 & 2006 & 2007 & 2008 & 2009\\
\midrule
\textbf{\cellcolor{gray!10}{n}} & \cellcolor{gray!10}{931} & \cellcolor{gray!10}{1,089} & \cellcolor{gray!10}{1,255} & \cellcolor{gray!10}{1,469} & \cellcolor{gray!10}{1,689}\\
\textbf{PctNoExpense} & 7.73 & 5.88 & 5.58 & 4.49 & 5.09\\
\textbf{\cellcolor{gray!10}{p25}} & \cellcolor{gray!10}{377} & \cellcolor{gray!10}{353} & \cellcolor{gray!10}{421} & \cellcolor{gray!10}{457} & \cellcolor{gray!10}{442}\\
\textbf{p50} & 965 & 857 & 1,019 & 1,072 & 1,000\\
\textbf{\cellcolor{gray!10}{p75}} & \cellcolor{gray!10}{2,293} & \cellcolor{gray!10}{1,995} & \cellcolor{gray!10}{2,353} & \cellcolor{gray!10}{2,415} & \cellcolor{gray!10}{2,381}\\
\addlinespace
\textbf{p90} & 6,331 & 4,743 & 6,129 & 6,873 & 6,665\\
\textbf{\cellcolor{gray!10}{p95}} & \cellcolor{gray!10}{13,567} & \cellcolor{gray!10}{10,440} & \cellcolor{gray!10}{12,530} & \cellcolor{gray!10}{12,824} & \cellcolor{gray!10}{15,095}\\
\textbf{p96} & 17,177 & 15,045 & 16,844 & 14,812 & 21,437\\
\textbf{\cellcolor{gray!10}{p97}} & \cellcolor{gray!10}{20,182} & \cellcolor{gray!10}{19,150} & \cellcolor{gray!10}{20,385} & \cellcolor{gray!10}{19,038} & \cellcolor{gray!10}{26,588}\\
\textbf{p98} & 34,044 & 26,450 & 28,116 & 24,707 & 38,919\\
\addlinespace
\textbf{\cellcolor{gray!10}{p99}} & \cellcolor{gray!10}{42,818} & \cellcolor{gray!10}{40,854} & \cellcolor{gray!10}{43,186} & \cellcolor{gray!10}{39,265} & \cellcolor{gray!10}{57,110}\\
\textbf{p995} & 75,095 & 68,072 & 54,982 & 61,353 & 70,828\\
\textbf{\cellcolor{gray!10}{p999}} & \cellcolor{gray!10}{98,802} & \cellcolor{gray!10}{108,815} & \cellcolor{gray!10}{106,887} & \cellcolor{gray!10}{132,189} & \cellcolor{gray!10}{165,958}\\
\textbf{max} & 105,272 & 168,989 & 144,749 & 153,098 & 201,089\\
\textbf{\cellcolor{gray!10}{mean}} & \cellcolor{gray!10}{3,397} & \cellcolor{gray!10}{2,973} & \cellcolor{gray!10}{3,259} & \cellcolor{gray!10}{3,311} & \cellcolor{gray!10}{3,791}\\
\addlinespace
\textbf{sd} & 9,557 & 9,741 & 8,983 & 9,333 & 11,983\\
\bottomrule
\end{tabular}
\label{tab:table3.16}
\end{table}
\endgroup
\textbf{n}: total of individuals in portfolio;\textbf{p(k)}: k-th sample percentile;\textbf{SD}: standard deviation

\medskip

% latex table generated in R 4 2.3 by append_kable_to_file function
% Current Date/Time: 2024-02-18 01:18:04
\begingroup
\renewcommand{\thetable}{3.17}
\begin{table}[H]
\centering
\caption{Descriptive Statistics Annual Expenses (Female) 61a65}
\centering
\begin{tabular}[t]{>{}l>{\raggedleft\arraybackslash}p{2.5cm}>{\raggedleft\arraybackslash}p{2.5cm}>{\raggedleft\arraybackslash}p{2.5cm}>{\raggedleft\arraybackslash}p{2.5cm}>{\raggedleft\arraybackslash}p{2.5cm}}
\toprule
Description & 2005 & 2006 & 2007 & 2008 & 2009\\
\midrule
\textbf{\cellcolor{gray!10}{n}} & \cellcolor{gray!10}{113} & \cellcolor{gray!10}{226} & \cellcolor{gray!10}{366} & \cellcolor{gray!10}{538} & \cellcolor{gray!10}{741}\\
\textbf{PctNoExpense} & 3.54 & 6.64 & 5.19 & 6.69 & 6.48\\
\textbf{\cellcolor{gray!10}{p25}} & \cellcolor{gray!10}{675} & \cellcolor{gray!10}{732} & \cellcolor{gray!10}{643} & \cellcolor{gray!10}{824} & \cellcolor{gray!10}{784}\\
\textbf{p50} & 1,250 & 1,415 & 1,509 & 1,803 & 1,732\\
\textbf{\cellcolor{gray!10}{p75}} & \cellcolor{gray!10}{2,954} & \cellcolor{gray!10}{2,864} & \cellcolor{gray!10}{3,455} & \cellcolor{gray!10}{4,261} & \cellcolor{gray!10}{3,444}\\
\addlinespace
\textbf{p90} & 6,080 & 5,454 & 8,467 & 9,869 & 7,437\\
\textbf{\cellcolor{gray!10}{p95}} & \cellcolor{gray!10}{10,928} & \cellcolor{gray!10}{8,319} & \cellcolor{gray!10}{17,779} & \cellcolor{gray!10}{20,360} & \cellcolor{gray!10}{15,702}\\
\textbf{p96} & 11,122 & 11,551 & 20,114 & 25,948 & 19,334\\
\textbf{\cellcolor{gray!10}{p97}} & \cellcolor{gray!10}{11,801} & \cellcolor{gray!10}{14,764} & \cellcolor{gray!10}{26,854} & \cellcolor{gray!10}{28,537} & \cellcolor{gray!10}{26,256}\\
\textbf{p98} & 12,380 & 16,414 & 33,172 & 37,319 & 30,266\\
\addlinespace
\textbf{\cellcolor{gray!10}{p99}} & \cellcolor{gray!10}{17,442} & \cellcolor{gray!10}{20,332} & \cellcolor{gray!10}{56,645} & \cellcolor{gray!10}{63,507} & \cellcolor{gray!10}{42,492}\\
\textbf{p995} & 19,103 & 44,430 & 60,399 & 71,529 & 126,540\\
\textbf{\cellcolor{gray!10}{p999}} & \cellcolor{gray!10}{20,255} & \cellcolor{gray!10}{56,734} & \cellcolor{gray!10}{150,422} & \cellcolor{gray!10}{132,337} & \cellcolor{gray!10}{240,232}\\
\textbf{max} & 20,543 & 59,669 & 195,519 & 135,830 & 274,612\\
\textbf{\cellcolor{gray!10}{mean}} & \cellcolor{gray!10}{2,603} & \cellcolor{gray!10}{2,898} & \cellcolor{gray!10}{4,505} & \cellcolor{gray!10}{5,116} & \cellcolor{gray!10}{4,905}\\
\addlinespace
\textbf{sd} & 3,544 & 5,850 & 13,086 & 12,251 & 18,070\\
\bottomrule
\end{tabular}
\label{tab:table3.17}
\end{table}
\endgroup
\textbf{n}: total of individuals in portfolio;\textbf{p(k)}: k-th sample percentile;\textbf{SD}: standard deviation

\medskip

% latex table generated in R 4 2.3 by append_kable_to_file function
% Current Date/Time: 2024-02-18 01:18:04
\begingroup
\renewcommand{\thetable}{3.18}
\begin{table}[H]
\centering
\caption{Descriptive Statistics Annual Expenses (Male) 61a65}
\centering
\begin{tabular}[t]{>{}l>{\raggedleft\arraybackslash}p{2.5cm}>{\raggedleft\arraybackslash}p{2.5cm}>{\raggedleft\arraybackslash}p{2.5cm}>{\raggedleft\arraybackslash}p{2.5cm}>{\raggedleft\arraybackslash}p{2.5cm}}
\toprule
Description & 2005 & 2006 & 2007 & 2008 & 2009\\
\midrule
\textbf{\cellcolor{gray!10}{n}} & \cellcolor{gray!10}{132} & \cellcolor{gray!10}{270} & \cellcolor{gray!10}{424} & \cellcolor{gray!10}{603} & \cellcolor{gray!10}{811}\\
\textbf{PctNoExpense} & 8.33 & 8.89 & 8.25 & 7.63 & 7.52\\
\textbf{\cellcolor{gray!10}{p25}} & \cellcolor{gray!10}{361} & \cellcolor{gray!10}{313} & \cellcolor{gray!10}{411} & \cellcolor{gray!10}{539} & \cellcolor{gray!10}{480}\\
\textbf{p50} & 744 & 940 & 1,026 & 1,283 & 1,133\\
\textbf{\cellcolor{gray!10}{p75}} & \cellcolor{gray!10}{2,198} & \cellcolor{gray!10}{2,243} & \cellcolor{gray!10}{2,449} & \cellcolor{gray!10}{3,129} & \cellcolor{gray!10}{2,733}\\
\addlinespace
\textbf{p90} & 4,427 & 5,697 & 5,275 & 8,083 & 7,854\\
\textbf{\cellcolor{gray!10}{p95}} & \cellcolor{gray!10}{7,887} & \cellcolor{gray!10}{14,443} & \cellcolor{gray!10}{11,139} & \cellcolor{gray!10}{25,074} & \cellcolor{gray!10}{19,228}\\
\textbf{p96} & 8,416 & 16,886 & 13,019 & 31,824 & 23,151\\
\textbf{\cellcolor{gray!10}{p97}} & \cellcolor{gray!10}{8,923} & \cellcolor{gray!10}{18,602} & \cellcolor{gray!10}{15,341} & \cellcolor{gray!10}{41,806} & \cellcolor{gray!10}{28,318}\\
\textbf{p98} & 10,875 & 38,905 & 24,083 & 49,420 & 35,593\\
\addlinespace
\textbf{\cellcolor{gray!10}{p99}} & \cellcolor{gray!10}{20,273} & \cellcolor{gray!10}{49,467} & \cellcolor{gray!10}{29,582} & \cellcolor{gray!10}{107,065} & \cellcolor{gray!10}{90,216}\\
\textbf{p995} & 26,916 & 53,310 & 59,040 & 152,434 & 112,752\\
\textbf{\cellcolor{gray!10}{p999}} & \cellcolor{gray!10}{32,392} & \cellcolor{gray!10}{106,560} & \cellcolor{gray!10}{197,939} & \cellcolor{gray!10}{229,425} & \cellcolor{gray!10}{229,212}\\
\textbf{max} & 33,761 & 123,830 & 266,119 & 237,957 & 320,963\\
\textbf{\cellcolor{gray!10}{mean}} & \cellcolor{gray!10}{2,021} & \cellcolor{gray!10}{3,500} & \cellcolor{gray!10}{3,398} & \cellcolor{gray!10}{5,923} & \cellcolor{gray!10}{4,872}\\
\addlinespace
\textbf{sd} & 4,053 & 10,600 & 14,966 & 20,614 & 18,301\\
\bottomrule
\end{tabular}
\label{tab:table3.18}
\end{table}
\endgroup
\textbf{n}: total of individuals in portfolio;\textbf{p(k)}: k-th sample percentile;\textbf{SD}: standard deviation

\medskip

\begin{landscape}
    \begin{figure}[H]
        \centering
        \includegraphics[scale=.9]{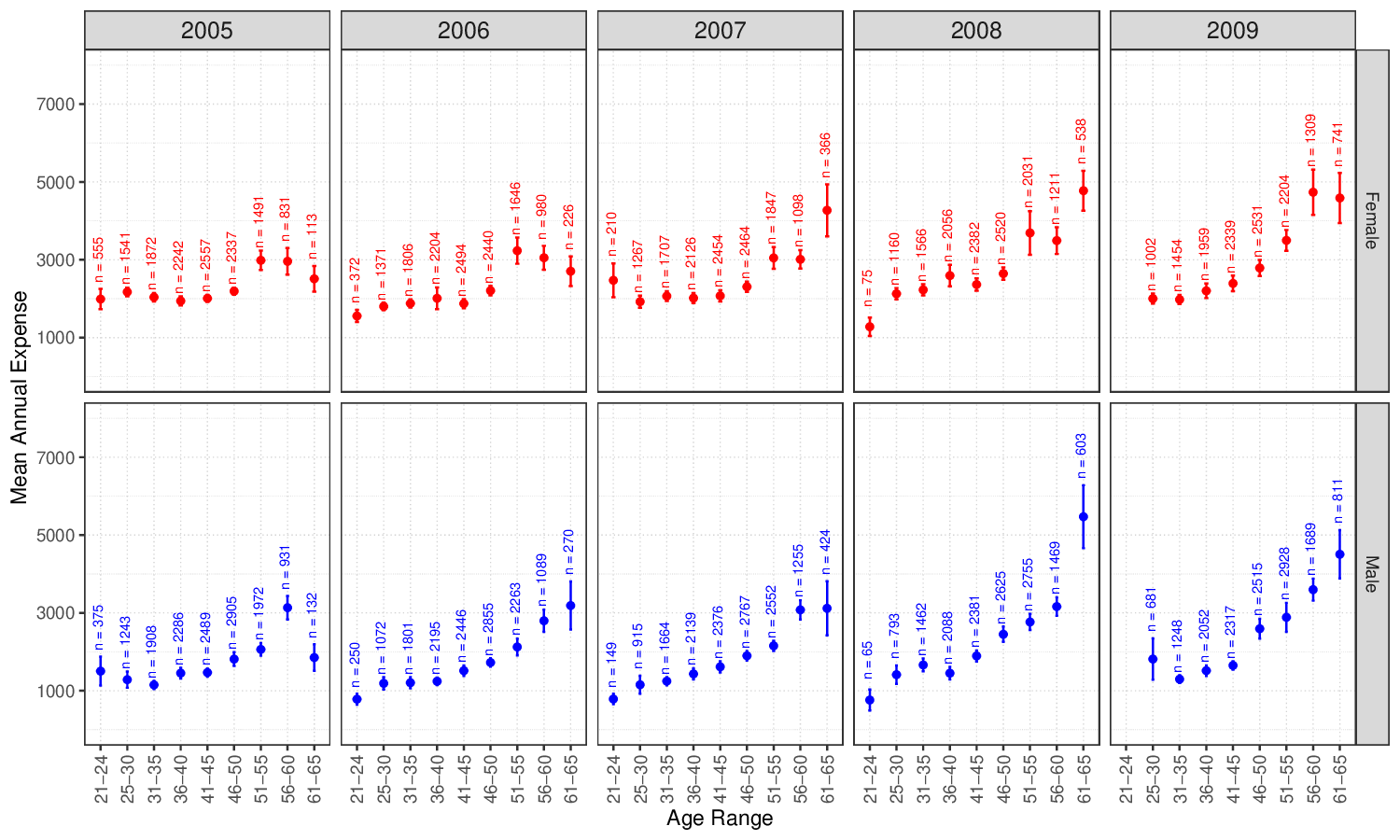}
        \caption{Profile plot of the mean annual expenses by year, for each sex and age range considering only individuals between 25 and 65 years-old which stayed on the portfolio during the whole five years period.}
        \label{fig:ci2005a2009_facet}  % 
    \end{figure}
\end{landscape}

\medskip

\begin{figure}[H]
    \centering
    \includegraphics[scale=.9]{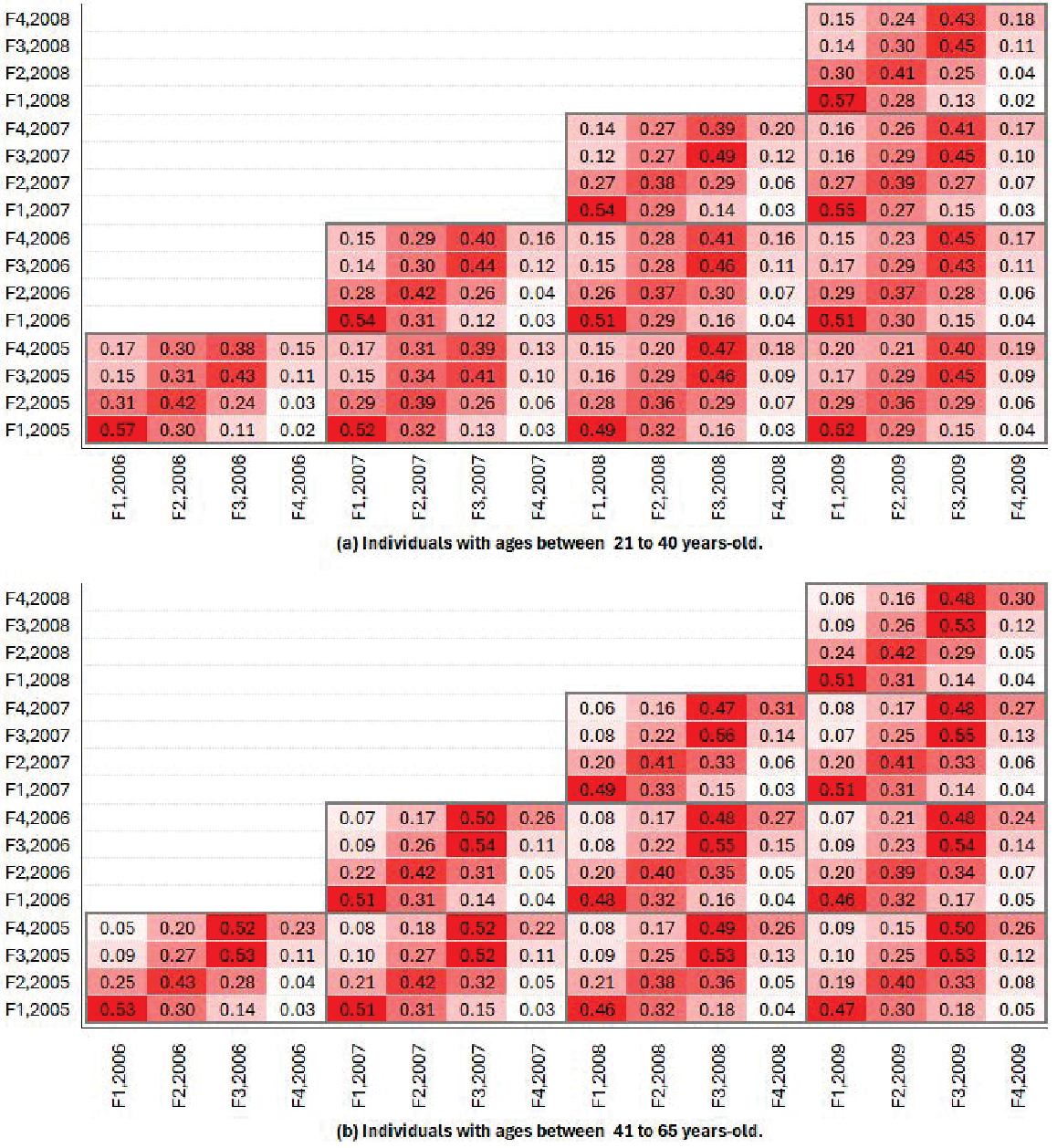}
    \caption{Transition matrices of expenses for each two years combination, i.e. from $2005$ to $2009$, considering individuals with ages between (a) 21 to 40 years-old, and (b) 41 to 65 years-old.}
    \label{fig:trmat_fig3_4}  % 
\end{figure}

\clearpage

\pagebreak

\section*{Appendix B - Markov Chains and HSA simulation}

\medskip

The Markov Chain approach proposed in this paper aims to predict the expense level of an individual in a given year by estimating the probability of his or her annual expenses being at every level given the expense level of previous years. Formally, considering the definition of $F_{j,i}$ presented in the section \ref{sec:markov}, the probability of an individual with annual expenses in $F_{k,i}$ at year $i$ having annual expenses in $F_{l,j}$ at year $j$ is estimated as:

\begin{equation}
    \hat{P}(F_{k,i},F_{l,j})=\dfrac{\text{Number of individuals in $F_{k,i}$ which are also in $F_{l,j}$}}{\text{Number of individuals in $F_{k,i}$}},\text{ with }j>i\label{eqn:appB_1}
\end{equation}

In Markov Chain Theory, $\hat{P}(F_{k,i},F_{l,j})$ is an estimator of the \textit{transition probabilities} between \textit{states} $F_{k,i}$ and $F_{l,j}$. Indeed, the ratio above is the proportion of individuals which were in $F_{k,i}$ at year $i$ and transitioned to $F_{l,j}$ at year $j$. Therefore, if we were to estimate the probability of an individual in $F_{k,i}$ to be in $F_{l,j}$ at year $j$ we could use the proportion $\hat{P}(F_{k,i},F_{l,j})$, which is a consistent estimator for such probability \cite{Anderson_Goodman1957}.

The approach above may also be applied to estimate the transition probabilities for individuals of a given sex and/or age range. This probability would allow to study the persistence of costs in distinct groups of individuals, as it is believed to behaviour differently for each age range and sex. The estimation of these probabilities would be done in the same manner as above, but the number of individuals considered in the ratio of $\hat{P}(F_{k,i},F_{l,j})$ would be that within the group of interest. Proceeding this way, we have a prediction for the probability of an individual, with given sex, age range and expense level in a previous year, to be in each expense level in the current year.

A transition matrix characterizes a Markov Chain, that is a sequence of random variables in which the distribution of the current random variable depends only on the value of the previous $k$ and is given by the probabilities of the transition matrix, in which $k$ is called the order of the Markov Chain. If $k = 1$ then the transition matrix is calculated as the ratio given in \eqref{eqn:appB_1} when $j = i + 1$. If $k = 2$, then the probability of being in a given state after visiting some in the past depends only on the last two states visited. In this case, the Markov Chain is generated by the transition probabilities estimated as

\begin{equation}
    \hat{P}(F_{k,i-2},F_{m,i-1};F_{l,i})=\dfrac{\text{Number of individuals in $F_{k,i-2}$ which are also in $F_{m,i-1}$ and $F_{l,i}$}}{\text{Number of individuals in $F_{k,i-2}$ which are also in $F_{m,i-1}$}}\label{eqn:appB_2}
\end{equation}

which refers to the transition to $F_{l,i}$ after being in $F_{k,i-2},F_{m,i-1}$ at the previous two years. These probabilities may also be estimated for a given group of individuals, considering the numbers in the ratio to be that within the group.

With a Markov Chain of order 2 we may estimate the probability of an individual with given sex and age range to be in each expense level as a function of the levels he or she was in the last two years. This is the method we use to predict the expense level of each life in the simulated portfolio. Observe that it diverges from the usual methods based on regression as we do not try to predict the exact value of the annual expense, but rather the expense level, so we do not have to assume a distribution for the expenses, nor a functional relation between the current year expenses and the independent variables sex, age range and previous two years expenses.

\medskip

\textbf{Transition Matrices}: In order to perform the first step of the simulation, we need to estimate a transition matrix between the expense levels. For this purpose, we suppose that the expense level follows a \textit{homogeneous} Markov Chain of order 2 \cite{Taylor_Karlin1998}. This means the transition from one state to another does not depend on the year, but only on the states. Formally, this means that

\begin{equation}
    P(F_{k,i-2},F_{m,i-1};F_{l,i})=P(F_{k,j-2},F_{m,j-1};F_{l,j})
\end{equation}

for any $i,j$, in which $P$ is the population transition probability, in contrast to the estimated transition probability $\hat{P}$. With this assumption, we may estimate the transition probabilities by that of the time period 2007 to 2009. Therefore, we estimate the transition from expense levels $F_k, F_m$ to $F_l$ as

\begin{equation}
    \hat{P}(F_k,F_m;F_l)=\hat{P}(F_{k,2007},F_{m,2008};F_{l,2009})
\end{equation}

in which we consider only the individuals within the respective sex and age range group in the ratios that define these transitions. This assumption is supported by the matrices in Figure \ref{fig:trmat_fig2}, where we see that those which compare years with the same distance are similar (these are the matrices in a same diagonal of Figure \ref{fig:trmat_fig2}).

Under this approach, we have 16 transition matrices, one for each combination of sex and age range (26-30, 31-35, 36-40, 41-45, 46-50, 51-55, 56-60, 61-65). Each matrix has 16 rows, one for each pair of expense levels in the current and last year, and 4 columns, one for each possible expense level in the next year. The entries of these matrices are transition probabilities from the state given by the pair to each one of the expense levels. To estimate the transitions, we consider that the age range of each individual in the dataset is the one which he or she was part of for the most number of months in the triennial 2007-2009.

\medskip

\textbf{First step}: The expense level of an individual next year is predicted based on his or her sex, age range and expense level last year and in the current year. This prediction is sampled from the estimated conditional distribution of the expense levels of given sex, age range and two last expense levels, which is the row associated to the last two expense levels of the transition matrix of the given sex and age range. We start this process at the initial values for 24 and 25 years-old to estimate the level at 26 years-old. We then iterate this process to estimate the level for the following ages: use the estimated values for 25 and 26 to predict 27 years-old; that of 26 and 27 to predict 28 years-old, and so on until the age of 65. Therefore, for each one of the 10,000 lives, we have a sequence of 41 levels corresponding to its predicted expense level for each work life year. Observe that, from age 27 on, the expenses history is predicted in the previous two years.

\medskip

\begin{figure}[H]
    \centering
    \includegraphics[scale=0.44]{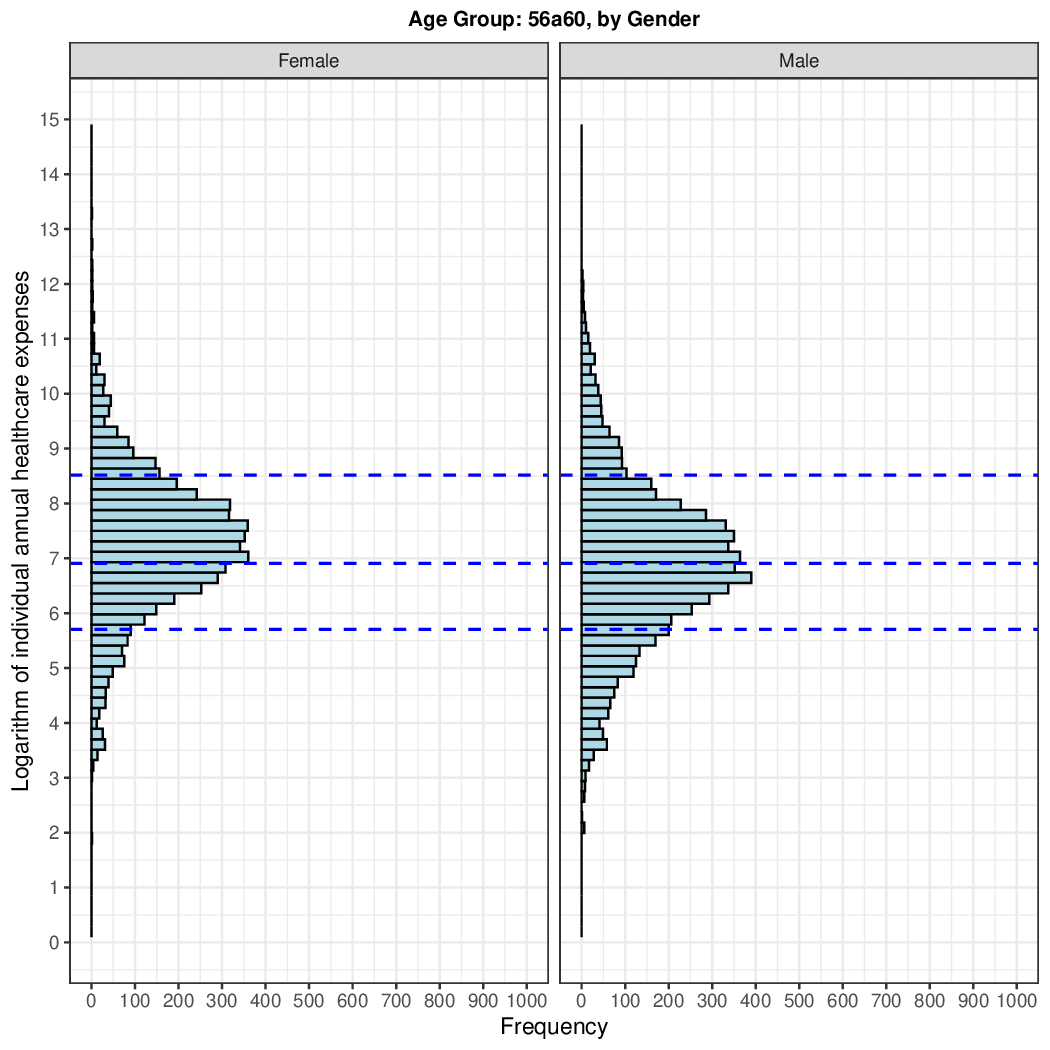}
    \centering
    \includegraphics[scale=0.44]{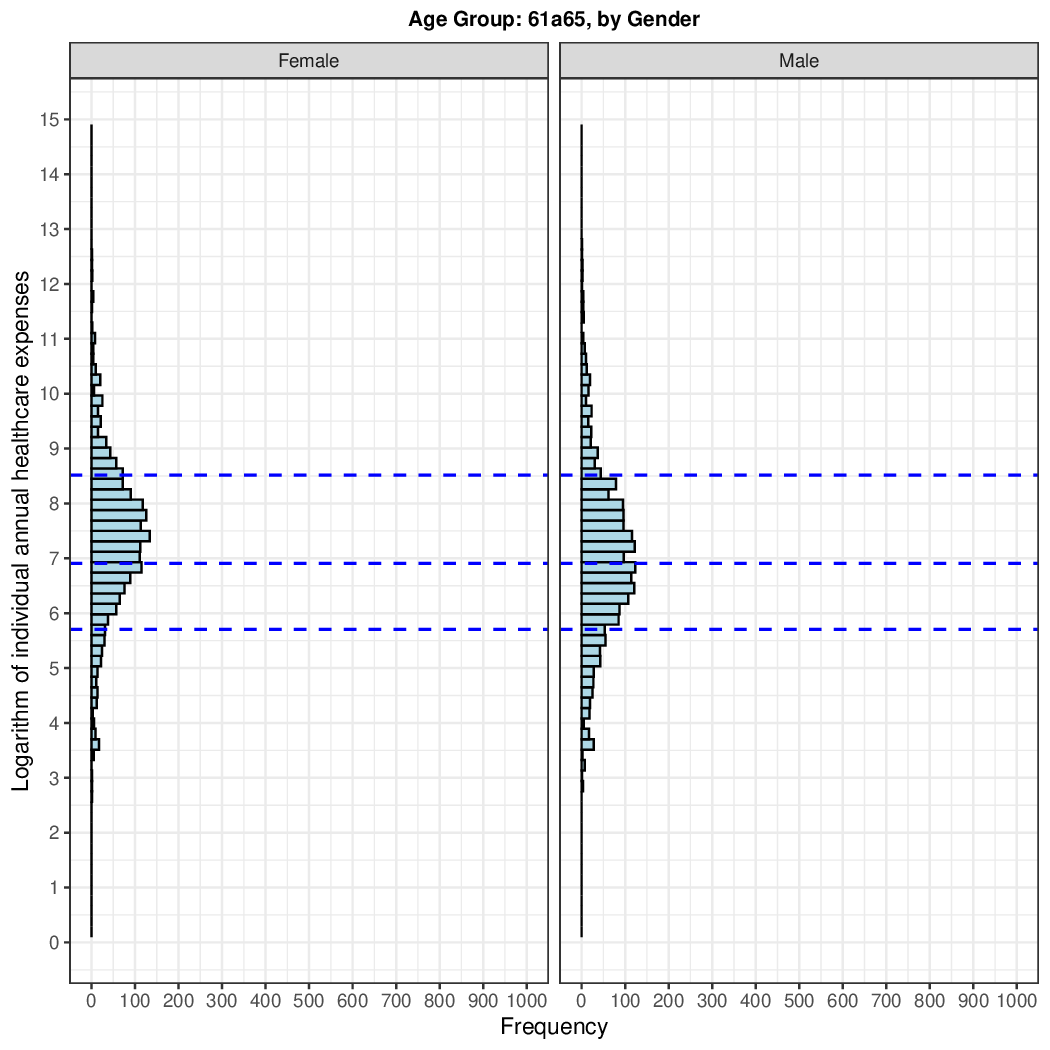}
    \caption{Empirical distributions of the logarithm of the annual healthcare expenses between 2005 and 2009. The distributions aggregate points from the five years. The dotted lines are respectively $\log(300)$, $\log(1000)$ and $\log(5000)$, which are the break points of the expense levels $F_{i,j}$.}
    \label{fig:empdistlbl}
\end{figure}

\textbf{Second step}: After we simulate the expense level of a life in a given year, we need to simulate the value to be withdrawn from its account to cover a healthcare expense in such level. This is done by sampling a point from the empirical distribution of the aggregated annual expenses of all years which contain only the individuals with the same sex and age range of the life. Within this empirical distribution, we sample one of the points inside the expense level predicted on step one, i.e., inside the respective dotted lines in Figure \ref{fig:empdistlbl}, which shows the aggregated empirical distributions of the logarithm of the annual healthcare expenses between 2005 and 2009 for each sex and age range.

\medskip

\begin{figure}[H]
    \centering
    \includegraphics[scale=0.44]{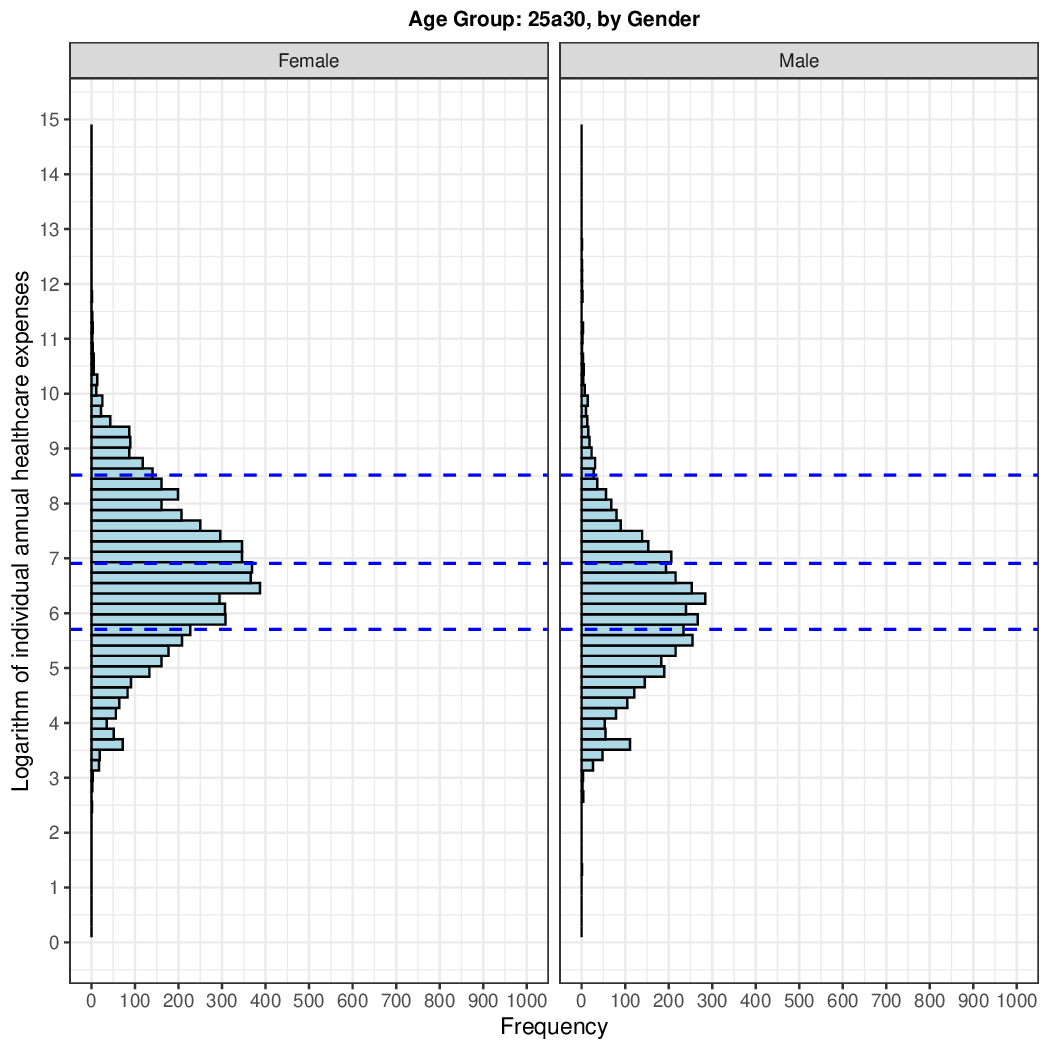}
    \centering
    \includegraphics[scale=0.44]{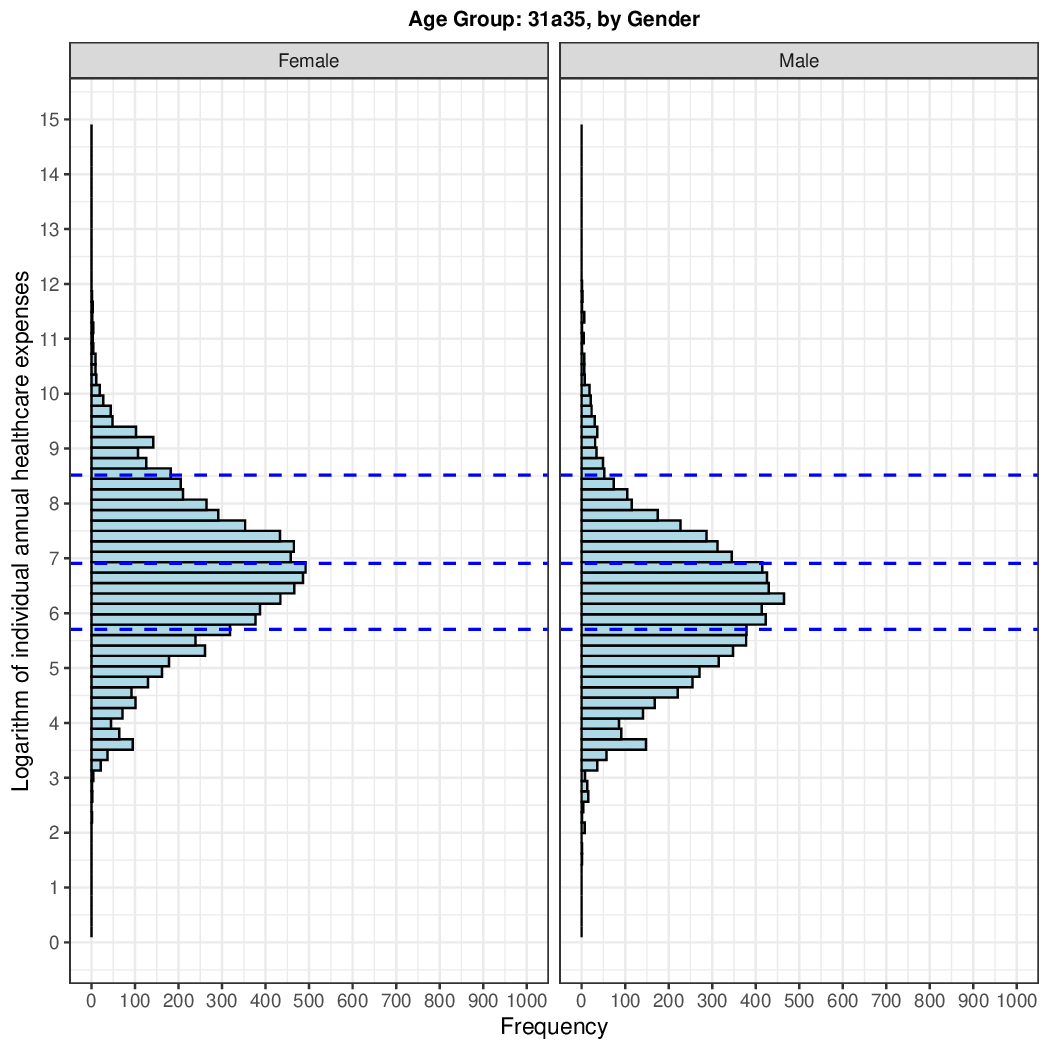}
\end{figure}

\begin{figure}[H]
    \centering
    \includegraphics[scale=0.44]{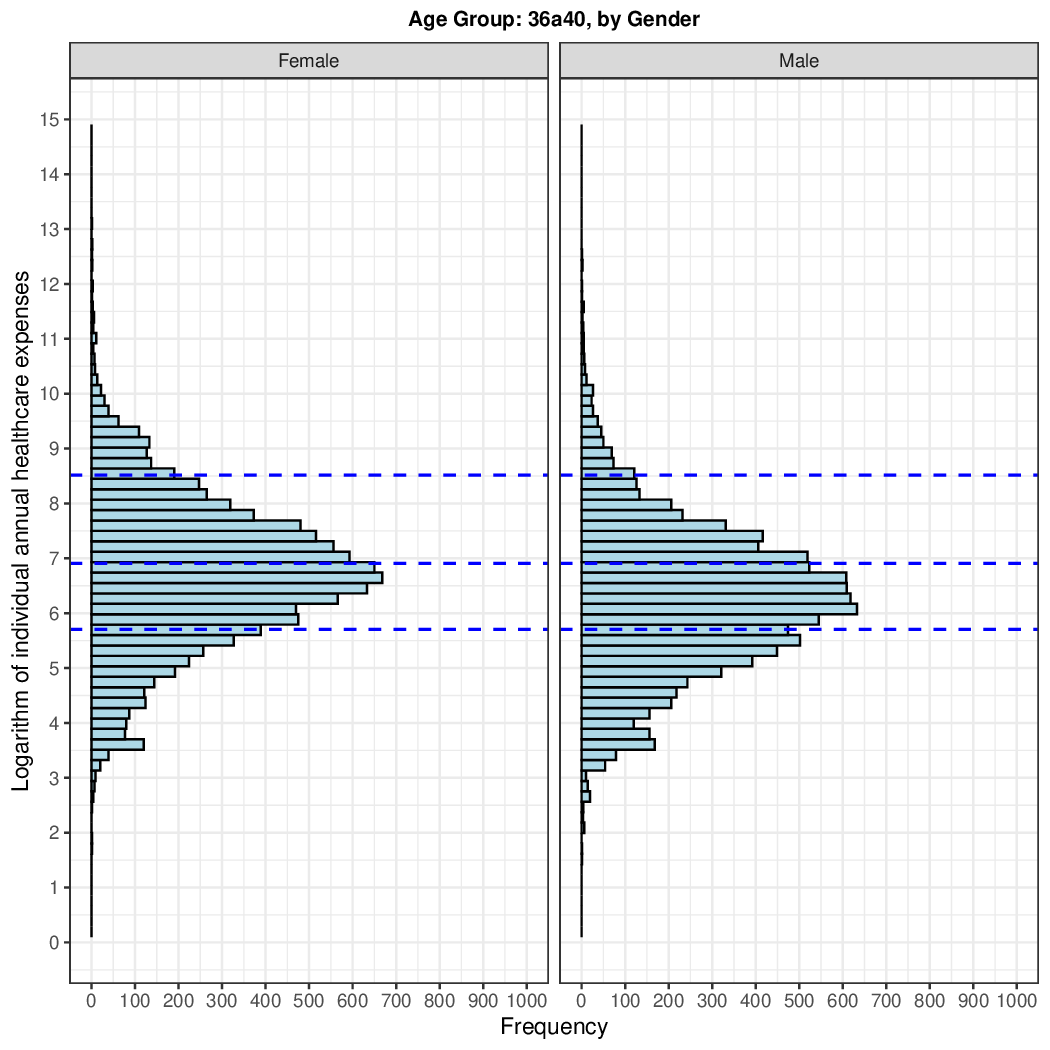}
    \centering
    \includegraphics[scale=0.44]{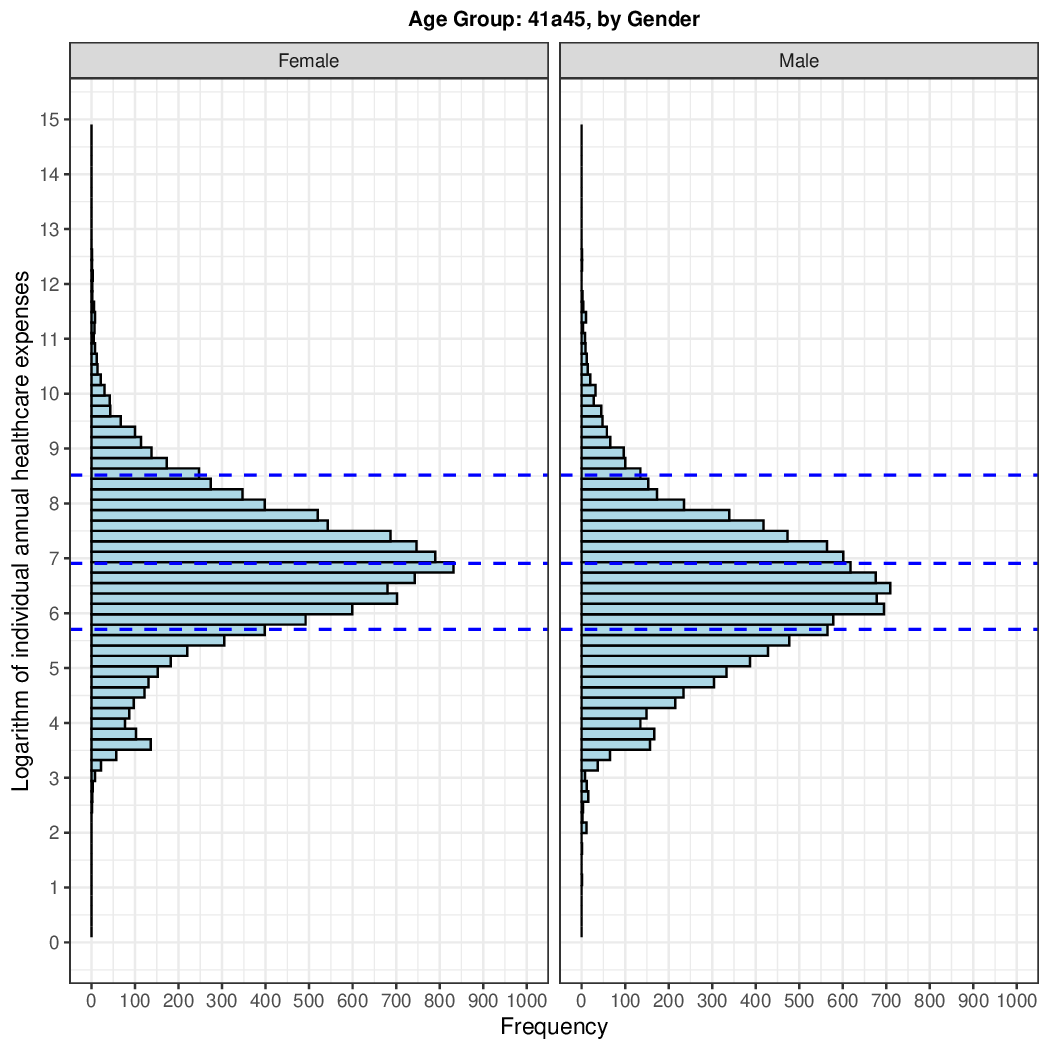}
\end{figure}

\begin{figure}[H]
    \centering
    \includegraphics[scale=0.44]{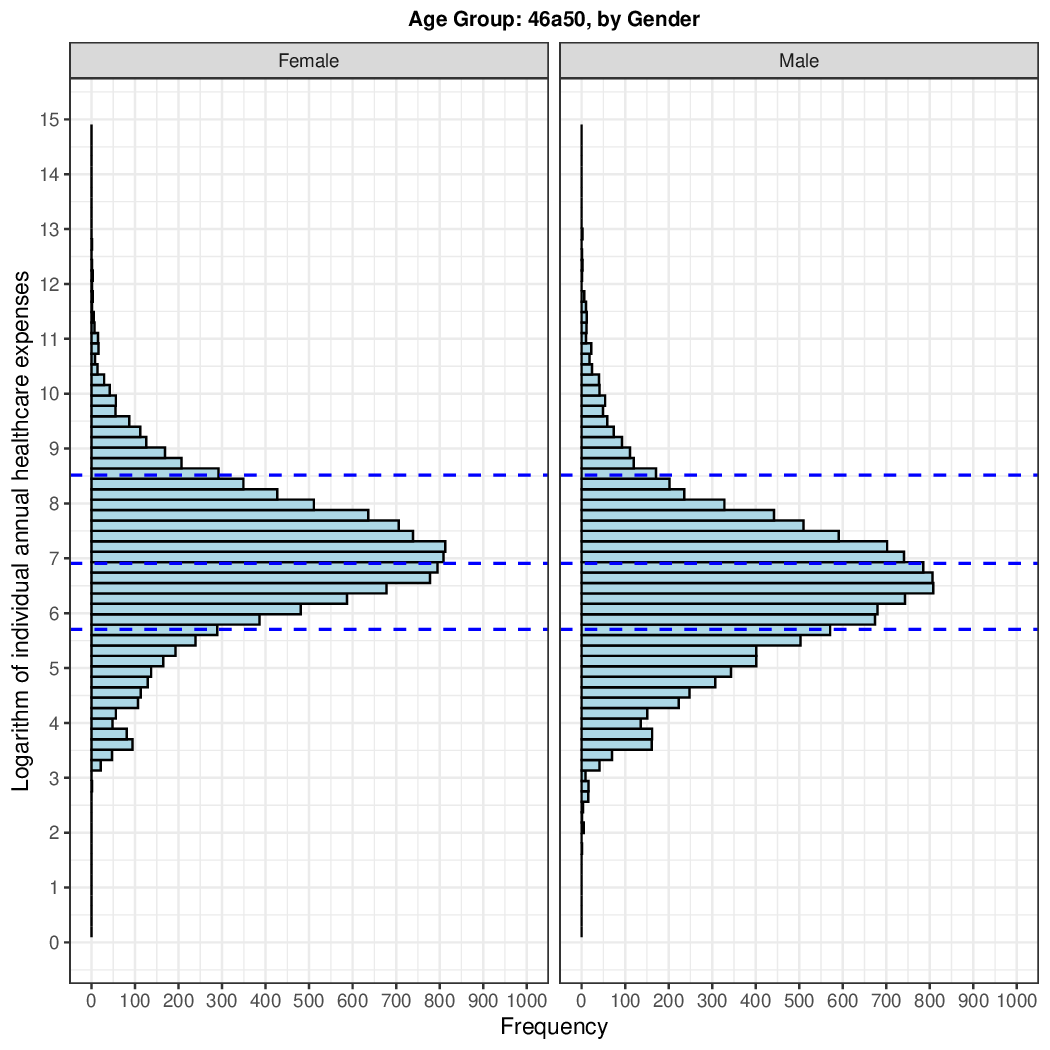}
    \centering
    \includegraphics[scale=0.44]{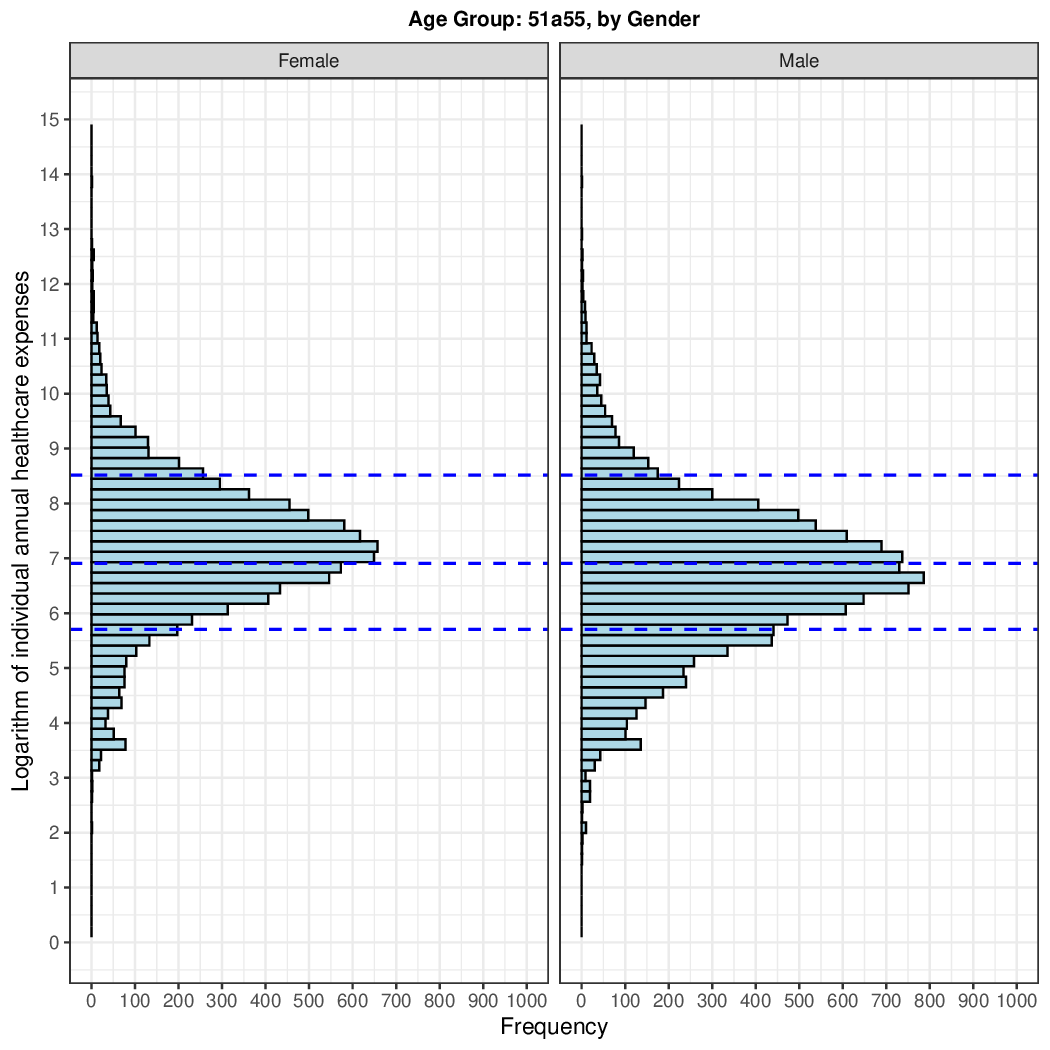}
\end{figure}

\medskip

In Table 4 we see the number of expenses in the dataset at each expense level by sex and age range. These percentages aggregate all expenses of the period, so they refer to the percentage of expenses, rather than the percentage of individuals, as each individual has five expenses, one for each year. In the dataset, 25\% of women’s expenses and 45\% of men’s in the age range 25-30 are in the first level (up to R\$ 300), percentages which decrease with age, attaining the values of 15\% and 23\% for women’s and men’s, respectively, at the age range 61-65. As the age increases, the same reduction is observed in the percentage of expenses in the second expense level (R\$ 300 - R\$ 1,000) for both sexes. On the other hand, the percentage of expenses in the third expense level (R\$ 1,000 - R\$ 5,000) increases with age and the difference between these percentages in age ranges 25-30 and 61-65 is 14 and 17 percentage points for women’s and men’s, respectively. The same is observed for the fourth expense level (greater than R\$ 5,000): 10\% of women’s expenses and 4\% of men’s in the age range 25-30 are greater than R\$ 5,000 in opposition to 16\% of women’s and 12\% of men’s in the age range 61-65. We see in the last three age ranges (after 51 years-old) that around 58\% to 62\% of women’s expenses, and 40\% to 48\% of men’s, are greater than R\$ 1,000.

\medskip

% latex table generated in R 4 2.3 by append_kable_to_file function
% Current Date/Time: 2023-12-30 00:13:33
\begingroup
\begin{table}[H]
\fontsize{9}{11}\selectfont
\begin{tabular}[t]{>{\centering\arraybackslash}p{1.5cm}>{\centering\arraybackslash}p{1.5cm}>{\centering\arraybackslash}p{1.5cm}>{\centering\arraybackslash}p{1.5cm}>{\centering\arraybackslash}p{1.5cm}>{\centering\arraybackslash}p{1.5cm}ccc}
\toprule
& & \textbf{Female} & & & & \textbf{Male}\\
Age
Range & {}[0-300] & (300-1000] & (1000-5000] & >5000 & {}[0-300] & (300-1000] & (1000-5000] & >5000\\
\midrule
\textbf{\cellcolor{gray!6}{25a30}} & \cellcolor{gray!6}{1,597} & \cellcolor{gray!6}{2,071} & \cellcolor{gray!6}{2,076} & \cellcolor{gray!6}{597} & \cellcolor{gray!6}{2,138} & \cellcolor{gray!6}{1,531} & \cellcolor{gray!6}{861} & \cellcolor{gray!6}{174}\\
\textbf{} & (25.2\%) & (32.7\%) & (32.7\%) & (9.4\%) & (45.5\%) & (32.5\%) & (18.3\%) & (3.7\%)\\
\textbf{\cellcolor{gray!6}{31a35}} & \cellcolor{gray!6}{2,091} & \cellcolor{gray!6}{2,741} & \cellcolor{gray!6}{2,790} & \cellcolor{gray!6}{783} & \cellcolor{gray!6}{3,374} & \cellcolor{gray!6}{2,690} & \cellcolor{gray!6}{1,703} & \cellcolor{gray!6}{316}\\
\textbf{} & (24.9\%) & (32.6\%) & (33.2\%) & (9.3\%) & (41.7\%) & (33.3\%) & (21.1\%) & (3.9\%)\\
\textbf{\cellcolor{gray!6}{36a40}} & \cellcolor{gray!6}{2,678} & \cellcolor{gray!6}{3,581} & \cellcolor{gray!6}{3,485} & \cellcolor{gray!6}{843} & \cellcolor{gray!6}{4,121} & \cellcolor{gray!6}{3,683} & \cellcolor{gray!6}{2,473} & \cellcolor{gray!6}{483}\\
\addlinespace
\textbf{} & (25.3\%) & (33.8\%) & (32.9\%) & (8.0\%) & (38.3\%) & (34.2\%) & (23.0\%) & (4.5\%)\\
\textbf{\cellcolor{gray!6}{41a45}} & \cellcolor{gray!6}{2,633} & \cellcolor{gray!6}{4,133} & \cellcolor{gray!6}{4,522} & \cellcolor{gray!6}{938} & \cellcolor{gray!6}{4,153} & \cellcolor{gray!6}{4,134} & \cellcolor{gray!6}{3,080} & \cellcolor{gray!6}{642}\\
\textbf{} & (21.5\%) & (33.8\%) & (37.0\%) & (7.7\%) & (34.6\%) & (34.4\%) & (25.6\%) & (5.3\%)\\
\textbf{\cellcolor{gray!6}{46a50}} & \cellcolor{gray!6}{2,198} & \cellcolor{gray!6}{3,765} & \cellcolor{gray!6}{5,194} & \cellcolor{gray!6}{1,135} & \cellcolor{gray!6}{4,201} & \cellcolor{gray!6}{4,703} & \cellcolor{gray!6}{3,892} & \cellcolor{gray!6}{871}\\
\textbf{} & (17.9\%) & (30.6\%) & (42.3\%) & (9.2\%) & (30.7\%) & (34.4\%) & (28.5\%) & (6.4\%)\\
\addlinespace
\textbf{\cellcolor{gray!6}{51a55}} & \cellcolor{gray!6}{1,354} & \cellcolor{gray!6}{2,545} & \cellcolor{gray!6}{4,258} & \cellcolor{gray!6}{1,062} & \cellcolor{gray!6}{3,265} & \cellcolor{gray!6}{4,117} & \cellcolor{gray!6}{4,166} & \cellcolor{gray!6}{922}\\
\textbf{} & (14.7\%) & (27.6\%) & (46.2\%) & (11.5\%) & (26.2\%) & (33.0\%) & (33.4\%) & (7.4\%)\\
\textbf{\cellcolor{gray!6}{56a60}} & \cellcolor{gray!6}{818} & \cellcolor{gray!6}{1,313} & \cellcolor{gray!6}{2,591} & \cellcolor{gray!6}{707} & \cellcolor{gray!6}{1,527} & \cellcolor{gray!6}{1,877} & \cellcolor{gray!6}{2,315} & \cellcolor{gray!6}{714}\\
\textbf{} & (15.1\%) & (24.2\%) & (47.7\%) & (13.0\%) & (23.7\%) & (29.2\%) & (36.0\%) & (11.1\%)\\
\textbf{\cellcolor{gray!6}{61a65}} & \cellcolor{gray!6}{309} & \cellcolor{gray!6}{442} & \cellcolor{gray!6}{917} & \cellcolor{gray!6}{316} & \cellcolor{gray!6}{524} & \cellcolor{gray!6}{646} & \cellcolor{gray!6}{796} & \cellcolor{gray!6}{274}\\
\addlinespace
\textbf{} & (15.6\%) & (22.3\%) & (46.2\%) & (15.9\%) & (23.4\%) & (28.8\%) & (35.5\%) & (12.2\%)\\
\bottomrule
\end{tabular}
\end{table}
\endgroup

\textbf{Table 4}. Percentage of expenses in each expense level by sex and age range. The expenses of all years are aggregated, so these percentages refer to the percentage of expenses, rather than the percentage of individuals, as each individual has five expenses, one for each year.

\twocolumn

\clearpage

\pagebreak

\Urlmuskip=0mu plus 1mu\relax

{\small\bibliographystyle{unsrtnat}
\bibliography{bibliography}}

\end{document}